\documentclass[useAMS,usenatbib,usegraphicx]{mn2e}
\usepackage[english]{babel}

\usepackage{fancyhdr}
\usepackage{amsfonts}
\usepackage{amsmath}
\usepackage{amssymb}
\usepackage{multicol}
\usepackage{layout}
\usepackage{graphicx}
\usepackage{multirow}
\usepackage{txfonts}

\usepackage{times}
\usepackage{natbib}

\newif\ifAMStwofonts
\AMStwofontstrue

\newcommand{\degree}{\ensuremath{^\circ}}

\graphicspath{{./fig/}}

\title[Cosmic web and substructure for spectra]{The importance of the cosmic web and halo substructure for power 
spectra}

\author[Pace et~al.]{Francesco Pace$^{1,2}$\thanks{francesco.pace@manchester.ac.uk}, Marc Manera$^{3,2}$, 
David J. Bacon$^{2}$, Robert Crittenden$^{2}$, Will J. Percival$^{2}$\\
$^1$ Jodrell Bank Centre for Astrophysics, School of Physics and Astronomy, The University of Manchester, Manchester, 
M13 9PL, U.K.\\
$^2$ Institute of Cosmology and Gravitation, University of Portsmouth, Dennis Sciama Building, Portsmouth, PO1 3FX, 
U.K.\\
$^3$ University College London, Gower Street, London, WC1E 6BT, U.K.}

\date{Accepted ?, Received ?; in original form ?}

\pagerange{\pageref{firstpage}--\pageref{lastpage}} \pubyear{2014}

\begin{document}
\label{firstpage}
\maketitle

\begin{abstract}
In this work we study the relevance of the cosmic web and substructures on the matter and lensing power spectra 
measured from halo mock catalogues extracted from the N-body simulations. 
Since N-body simulations are computationally expensive, it is common to use faster methods that approximate the dark 
matter field as a set of halos. 
In this approximation, we replace mass concentrations in N-body simulations by a spherically symmetric 
Navarro-Frenk-White halo density profile. 
We also consider the full mass field as the sum of two distinct fields: dark matter halos 
($M>9\times 10^{12}~M_{\odot}$/h) and particles not included into halos. 
Mock halos reproduce well the matter power spectrum, but underestimate the lensing power spectrum on large and small 
scales. 
For sources at $z_{\rm s}=1$ the lensing power spectrum is underestimated by up to 40\% at $\ell\approx 10^4$ with 
respect to the simulated halos. 
The large scale effect can be alleviated by combining the mock catalogue with the dark matter distribution outside 
the halos. 
In addition, to evaluate the contribution of substructures we have smeared out the intra-halo substructures in a 
N-body simulation while keeping the halo density profiles unchanged. 
For the matter power spectrum the effect of this smoothing is only of the order of 5\%, but for lensing substructures 
are much more important: for $\ell\approx 10^4$ the internal structures contribute 30\% of the total spectrum. 
These findings have important implications in the way mock catalogues have to be created, suggesting that some 
approximate methods currently used for galaxy surveys will be inadequate for future weak lensing surveys.
\end{abstract}

\begin{keywords}
 Gravitational lensing: weak - methods: analytical, numerical, statistical - cosmology: theory, large-scale structure 
 of Universe
\end{keywords}

\section{Introduction}
One of the most important ways to study the growth of cosmological structure and infer the nature of gravity is to 
use gravitational lensing. Lensing arises from the deflection of light rays due to structures along the line-of-sight 
between the source and the observer. 
In particular, weak gravitational lensing measures the shear effect produced by the large scale structures in the 
Universe on the background galaxies, making it an effective tool to evaluate the total matter in the universe. 
Weak lensing observations are steadily improving 
\citep{Kilbinger2013,Simpson2013,VanWaerbeke2013,Fu2014,Kitching2014} 
and on-going and future surveys such as KIDS\footnote{http://kids.strw.leidenuniv.nl}, 
DES\footnote{http://www.darkenergysurvey.org}, Pan-STARRS\footnote{http://pan-starrs.ifa.hawaii.edu}, 
LSST\footnote{http://www.lsst.org} and Euclid\footnote{http://www.ias.u-psud.fr/imEuclid} 
\citep{Laureijs2011,Amendola2013} will be able to probe the growth of structures and the expansion history to a very 
high precision. 
Such a precise determination of the growth of structures will provide constraints on the underlying cosmological model 
and probe whether General Relativity is a good description of the Universe (with cosmological constant or some form of 
dark energy) or whether modifications of gravity are required.

Theoretical expectations of weak lensing on small scales are often based on raytracing through simulations 
\citep{Pace2007.1,Hilbert2009,Fedeli2011,Hartlap2011,Pace2011,Pace2015}, where light rays are shot from the 
observer to distant sources through the matter distribution of an N-body simulation. These simulations are 
useful to probe lensing statistics for different cosmological models and highlight differences in the non-linear 
regime that are usually not accessible via analytical techniques. 
Many lensing studies have focused on the properties of the probability distribution function (PDF) of the effective 
convergence and shear 
\citep{Munshi2000,Taruya2002,Valageas2000a,Valageas2000b,Menard2003,Valageas2004a,Valageas2004b,Takahashi2011}, 
on how to model it numerically \citep{Kainulainen2009,Kainulainen2011,Marra2013}, how to combine it with other 
cosmological probes \citep {Song2011,Munshi2014a,Munshi2014b} and on tomography \citep{Munshi2014c}.

Much attention is also devoted to the study of lensing properties of a given mass density profile for dark matter 
halos \citep{Bartelmann1996,Oguri2011,Retana-Montenegro2011,Retana-Montenegro2012,Retana-Montenegro2012a} 
as according to N-body simulations, halo density profiles can be parametrised with a universal function 
\citep{Navarro1996,Navarro1997,Gao2008}. In addition, the halo density profile is one of the main ingredients for 
the halo model \citep{Scherrer1991,Seljak2000,Cooray2002,Sheth2003}; halo model applications for lensing include 
calculating the lensing power spectrum \citep{Cooray2000a,Takada2003c,Kainulainen2011}, the influence of 
substructures \citep{Giocoli2008,Giocoli2010}, and the modelling of the intrinsic alignments \citep{Schneider2010} 
and their effects on the lensing power spectrum \citep{Ciarlariello2014}. 
Conversely, lensing can be used to constrain elements of the halo model \citep{Cacciato2012,Velander2014}.

Halo modelling also plays an important role in creating fast realisations of large scale structure that can play an 
important role in modelling observations of large scale structure and particularly their covariances.  
Many different techniques have been put forward exploiting halo modelling. 
\cite{Coles1991} and \cite{Cole2005} used the log-normal model to generate mock catalogues; \cite{Chuang2015} 
used an approach based on the Zel'dovich approximation to create mock catalogues that can accurately reproduce the 
one-point, two-point, and three-point statistics; 
\cite{Scoccimarro2002} and \cite{Manera2013} used the 2LPT formalism to create matter density fields from which halo 
catalogues were extracted (PTHalos). \cite{delaTorre2013} instead used a method based on the sampling of the mass 
function to create halos under the limit resolution of the halos; 
\cite{Tassev2013,Tassev2015} used a method based on Lagrangian Perturbation Theory (LPT) and similar methods were 
used by \citep{Monaco2002a,Monaco2002b,Taffoni2002,Monaco2013}. 
A compilation of some of the fast methods for generating mock catalogues can be found in \cite{Chuang2014}.
Other techniques, based on remapping the halos from a simulation with a given cosmology to catalogues with a different 
cosmological background, have also been developed \citep{Mead2013}. 
Finally, methods to generate mock catalogues on a spherical surface by using a real distribution were also used for 
the 2MASS Redshift Survey (2MRS) catalogue \citep{DeDomenico2012}.\\

Halo model techniques have been tuned to reproduce galaxy clustering surveys, where they are exploited to calculate 
covariances of observations. 
It is important to understand whether they are similarly accurate for lensing surveys, as they can potentially be 
used to calculate not only lensing covariances, but also potential cross correlations and covariances for joint 
lensing and galaxy observations. Here we attempt to answer this question.

To do this, we create a convergence field using the halo catalogue extracted with a Friend-of-Friend (FoF) technique 
from an N-body simulation and compare its power spectrum with the convergence spectrum from the original N-body 
particles. We assume that the halos are well described by a Navarro-Frenk-White (NFW) density profile. 
In this way we can quantify the accuracy with which the halo catalogue reproduces the numerical simulation results, 
and on which scales we can rely on analytical methods.

We also study the importance of substructures in the simulated halos. To do so, we randomly rotate the particles in a 
halo while keeping them always at the same distance from the centre. In this way coherent subhalos are erased while 
the density profile of the halo remains unchanged. This creates numerical halos more similar to the analytical ones, 
since the NFW profile doesn't include the presence of substructures. 

Finally we explore how well fast halo catalogue techniques are able to reproduce full N-body simulations and the halo 
catalogues extracted from them. 
In particular, we focused on the PTHalos technique, based on realisations of the 2LPT field \citep{Manera2013} and 
compare its results with N-body simulations.

We begin in Sect.~\ref{sect:N-HM} with a brief description of the N-body simulations, and the implementation of the 
halo modelling, based on an NFW profile. 
In the following section, we compare the full N-body statistics to those derived from the halo catalogues, focusing 
in particular on the matter power spectrum (Sect.~\ref{sect:Nbody_Pk}) and the lensing convergence spectrum 
(Sect.~\ref{sect:Nbody_Pl}) and explore the impact of halo substructure. In Sect.~\ref{sect:2LPT} we investigate the 
effectiveness of a fast halo method, PTHalos, for calculating these statistics. 
Finally in Section~\ref{sect:CBS} we study the effectiveness of the optimal weight on lensing statistics, when only 
mock catalogues for halos are available, before drawing conclusions in Section~\ref{sect:conclusions}.

\upshape

\section{Simulations and the halo model}\label{sect:N-HM}
The aim of this work is to compare results from N-body simulations with mock catalogues from a statistical point of 
view, focusing in particular on the matter and lensing power spectra. 
We begin by describing the N-body simulations that will be the basis of the halo catalogues and the methods we use to 
approximate them using a halo model approach.

\subsection{N-body simulations} \label{sect:Nbodysim}
We use the LasDamas\footnote{http://lss.phy.vanderbilt.edu/lasdamas} N-body simulations 
\citep{McBride2009,McBride2011}, created with the publicly available code GADGET-II \citep{Springel2005}. 
In particular we use the suite of realizations called Oriana; the Oriana suite, an ensemble of 40 independent 
realizations of the dark matter distribution in the Universe, studies the dark matter particles evolution in a box of 
2.4~Gpc/h comoving filled with $1280^3$ particles. The mass resolution is $4.573\times10^{11}~M_{\odot}/h$ and the 
corresponding softening length is 53~kpc/h comoving.
Initial conditions were created at $z=49$ using the 2LPT formalism in order to take into account very early 
non-linearities and therefore to be more accurate than the usual Zel'dovich approximation \citep{Crocce2006}. The 
initial linear power spectrum was created using the CMBFAST code \citep{Seljak1996}. 
In Table~\ref{tab:Oriana} we show the cosmological parameters adopted for the Oriana simulations.

\begin{table}
 \centering
 \begin{tabular}{ll}
  \hline
  \hline
  Cosmological Parameter & Value\\
  \hline
  Baryon density $\Omega_{\rm b}$ & 0.04 \\
  Matter density $\Omega_{\rm m}$ & 0.25 \\
  Cosmological constant density $\Omega_{\Lambda}$ & 0.75 \\
  Hubble parameter $h$ & 0.7 \\
  Spectral index $n_{\rm s}$ & 1.0 \\
  Power spectrum normalization $\sigma_8$ & 0.8 \\
  \hline
  \hline
 \end{tabular}
  \caption[Parameters]{Cosmological parameters adopted for the N-body simulation Oriana.}
   \label{tab:Oriana}
 \begin{flushleft}
  \vspace{-0.5cm}
  {\small}
 \end{flushleft}
\end{table}

Within the simulations, halos are identified using a parallel Friends-of-Friends (FoF) algorithm \citep{Davis1985} 
implemented within the Ntropy framework \citep{Gardner2007a,Gardner2007b}. The linking length adopted is 0.2 times 
the mean inter-particle distance.

\begin{figure}
 \includegraphics[width=0.3\textwidth,angle=-90]{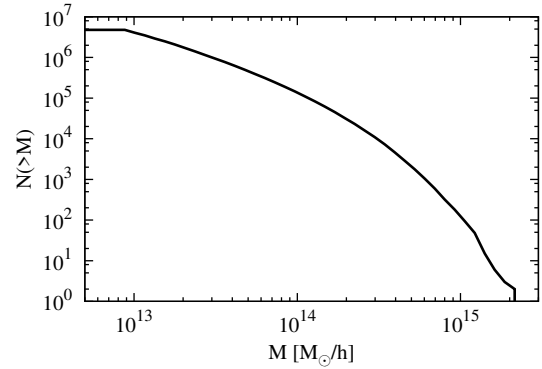}
 \caption{The number of objects above a given mass threshold $M$ for the Oriana simulation at $z=0.52$.}
 \label{fig:mf}
\end{figure}

In Fig.~\ref{fig:mf} we show the mass function of the Oriana simulation used in this work, evaluated from the FoF 
catalogue. We show the number of objects above a given mass. A halo is defined as an object with at least 20 
particles linked together, therefore the minimum mass probed is $M\approx 9\times 10^{12}~M_{\odot}/h$. As we show 
below, the sampled catalogue does not contain all of the galactic mass objects that might be contributing to the 
matter or lensing signal.

Below, we also explore whether halo locations and masses derived using spherical overdensity (SO) techniques differ 
significantly from the Friends-of-Friends method. However the halos are identified, all information about the 
internal structure of the halos is lost once the mock halo catalogues are constructed. One main goal is to quantify 
whether this simplification makes it impossible to recover the density and lensing spectra.

\subsection{The halo model approach}\label{sect:HM}
To analytically model the simulated halos, we describe them as spherically symmetric objects characterised by a 
Navarro-Frenk-White (NFW) density profile \citep{Navarro1996,Navarro1997}, whose functional form is:
\begin{equation}\label{eqn:NFW}
 \rho_{\rm NFW}(r)=\frac{\rho_{\rm s}}{(r/r_{\rm s})(1+r/r_{\rm s})^2}\;,
\end{equation}
where $\rho_{\rm s}=\Delta\rho_{\rm c}$ and $\rho_{\rm c}$ is the critical density and $\Delta=200$. The density 
$\rho_{\rm s}$ is related to the virial mass of the halo, $M_{\rm vir}$. 
The over-density is sometimes referred to the background comoving density. 
The quantity $r_{\rm s}$ is the scale radius and it is related to the virial radius (the radius of the halo) via the 
concentration parameter $c$, $r_{\rm s}=r_{\rm vir}/c$. In particular, the relation between $\rho_{\rm s}$ and 
$M_{\rm vir}$ is
\begin{equation}
 \rho_{\rm s}=\frac{M_{\rm vir}}{4\pi r_{\rm s}^3m(c)}\;,
\end{equation}
and $m(c)$ is the dimensionless mass of the NFW halo:
\begin{equation}
 m(c)=\int_{0}^{c}\frac{x}{(1+x)^2}dx=\ln(1+c)-\frac{c}{1+c}\;.
\end{equation}

The only remaining unknown quantity is the concentration parameter $c$. 
Many studies in the literature 
\citep{Navarro1997,Bullock2001,Eke2001,Neto2007,Duffy2008,Gao2008,Maccio2008,MunozCuartas2011,Prada2012,
Bhattacharya2013,Ludlow2013} have been devoted to the study of the dependence of the concentration parameter on the 
mass and on the redshift of the halo. Because low mass objects are formed earlier, when the average density of the 
universe was higher, they tend to have higher concentrations than higher mass objects. 
In this work we adopt the original concentration-mass relation established by \cite{Navarro1997}; however, the exact 
recipe adopted to evaluate the concentration does not greatly alter our results. 
Differences of order of a few percent appear in the convergence power spectrum on non-linear scales, with the 
prescription by \cite{Navarro1997} giving more power than the \cite{Bullock2001} and \cite{Eke2001} recipe. 

A useful feature of the NFW density profile is that many lensing quantities, and in particular the convergence, have 
an analytical expression \citep[see][]{Bartelmann1996,Wright2000}. 
The convergence is defined as the integral, along the line of sight, of the density profile scaled by the critical 
density and weighted by a function that depends on the relative distance of the lensing system components (observer, 
lens and source). 
Defining the projected matter density $\Sigma(\vec{\theta})=\int_{0}^{\infty}\rho(\vec{\theta},z)dz$ and the critical 
density $\Sigma_{\rm c}\equiv\frac{c^2}{4\pi G}\frac{D_{\rm s}}{D_{\rm d}D_{\rm ds}}$, the effective convergence is 
therefore $\kappa(\vec{\theta})\equiv\Sigma(\vec{\theta})/\Sigma_{\rm crit}$. 
Here, $c$ and $G$ represent the speed of light and the gravitational constant, while the inverse of the lensing 
efficiency in the definition of the critical density is given by the combination of three angular-diameter distances, 
$D_{\rm s}$, $D_{\rm d}$ and $D_{\rm ds}$, representing the distance between the observer and the source, the 
observer and the lens and the lens and the source, respectively. 
Finally, the vector $\vec{\theta}$ represents the angular position on the lens plane.

The radial dependence of the surface mass density (justified by the spherical symmetry) is
\begin{equation}\label{eqn:kappaNFW}
 \Sigma(x)=\left\{
 \begin{array}{lr}
  \frac{2r_{\rm s}\rho_{\rm s}}{\left(x^2-1\right)}\left[1-\frac{2}{\sqrt{1-x^2}}\rm{arctanh}\sqrt{\frac{1-x}{1+x}}
  \right] & x<1\\
  \frac{2r_{\rm s}\rho_{\rm s}}{3} & x=1\\
  \frac{2r_{\rm s}\rho_{\rm s}}{\left(x^2-1\right)}\left[1-\frac{2}{\sqrt{x^2-1}}\rm{arctan}\sqrt{\frac{x-1}{1+x}}
  \right] & x>1
 \end{array}
 \right.
\end{equation}
where we have defined $x\equiv r/r_{\rm s}$.
Note that all these equations imply a smooth inner profile, where substructures (density peaks inside the halos) do 
not play any role. For the comparison between the lensing signal of the numerical and analytical halos we will rely 
on Equation~\ref{eqn:kappaNFW} and we will show its accuracy and limitations for both N-body (see 
Sect.~\ref{sect:Nbody_Pl}) and 2LPT simulations (see Sect.~\ref{sect:2LPT_Pl}).

\section{Reproducing the N-body results}\label{sect:Nbody}
In this section we study the statistical properties of the matter distribution in the simulated box at $z=0.52$, 
focussing on the matter power spectrum (see Sect.~\ref{sect:Nbody_Pk}) and the lensing signal (see 
Sect.~\ref{sect:Nbody_Pl}).

We use a number of techniques to explore the impact of the substructure, and in particular how the halo catalogues 
perform in reproducing the density statistics. 
We directly compare the full N-body results to the FoF halo catalogues; to get comparable results, the halo catalogue 
must be rescaled appropriately as explained later so that its average density is the same as the N-body. 
We also divide the particles in the simulation into two different groups: those particles in halos (PIH) that, 
according to the FoF algorithm, build up the halos, and the particles outside the halos (POH) that contain the 
remaining particles. 
In terms of the particle numbers, it follows that $N_{\rm tot}=N_{\rm PIH}+N_{\rm POH}$. 
This division also allows us to directly test the impact of treating the halos as spheres.

To further probe the effects of halo substructure, we consider the impact of randomly rotating the positions of the 
PIH around the associated halo centres. We do this either by coherently rotating the particle positions, such that 
halo substructure is preserved, or by randomly rotating the individual particles while preserving their distances to 
the halo centres. In the first case, the power in the internal structures should remain unchanged, while the 
cross-correlations between the substructure and the cosmic web could be suppressed.  In the second case, the 
substructures are erased and the halos become effectively spherical, but their radial distributions are not fixed to 
a particular profile, as in the NFW halo catalogue.

\subsection{3D matter power spectrum}\label{sect:Nbody_Pk}
The 3D matter power spectrum represents an important statistical tool to probe the growth of fluctuations and the 
underlying cosmological model. 
To evaluate the matter power spectrum using the particle distribution in the simulated box, we divide the box volume 
into 512$^3$ voxels and we assign each dark matter particle to the corresponding voxel using a Nearest Grid Point 
(NGP) assignment scheme \citep{Hockney1988}. Once the discretized volume is created, we Fourier transform it and 
multiply it by its complex conjugate, according to the definition of the matter power spectrum
\begin{equation}\label{eqn:3D_spectrum}
 \langle\delta(\vec{k})\delta^{\ast}(\vec{k}^{\prime})\rangle=
 (2\pi)^3\delta^{(3)}(\vec{k}-\vec{k}^{\prime})P_{\delta}(\vec{k})\;.
\end{equation}
In the previous equation $\delta^{(3)}(\vec{k}-\vec{k}^{\prime})$ represents the three-dimensional Dirac's delta and
$\delta=(\rho-\bar{\rho})/\bar{\rho}$ the matter overdensity. 
For our analysis, the Nyquist frequency is $k_{\rm Nyq}\approx 0.67$~h/Mpc.

\begin{figure}
 \includegraphics[width=0.3\textwidth,angle=-90]{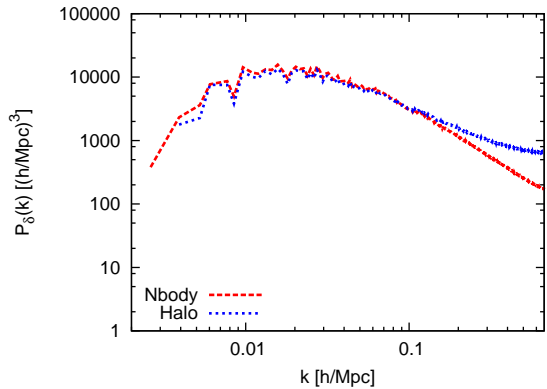}
 \caption{Matter power spectrum for the N-body simulation. The red dashed line shows the spectrum obtained from all 
 the particles in the simulation box while the blue short-dashed curve shows the rescaled mass-weighted spectrum of 
 the FoF halos.}
 \label{fig:nbody_matter_halo}
\end{figure}

A similar procedure is carried out to evaluate the mass-weighted matter power spectrum of the FoF halos where we 
assign the mass of each halo to the corresponding voxel determined according to the position of the halo as given in 
the catalogue. The comparison between the matter power spectrum from N-body and the mass-weighed spectrum is shown in 
Fig.~\ref{fig:nbody_matter_halo}. We also evaluate the shot-noise term due to the finite number of halos in the 
catalogues. The halo shot noise is evaluated as $P(k)=1/\bar{n}$, where $\bar{n}$ is the number density of the FoF 
halos. The shot-noise due to the finite number of particles in the simulated box is negligible.

The two spectra agree very well on large scales where the linear regime is still valid; this is obtained with an 
appropriate amplitude rescaling of the halos power spectra to take into account the different effective matter 
densities of the two fields (multiplying the halo spectrum by the factor $\Omega_{\rm m}/\Omega_{\rm halo}$ squared). 
Obviously the agreement is limited only to a restricted range of frequencies $k\lesssim 0.1$~h/Mpc, corresponding to 
the modes where the non-linear matter power spectrum starts to deviate from the linear prediction. For higher modes, 
we notice a flattening in the halo matter power spectrum, due to the shot noise term that dominates for high $k$.

\begin{figure}
 \includegraphics[width=0.3\textwidth,angle=-90]{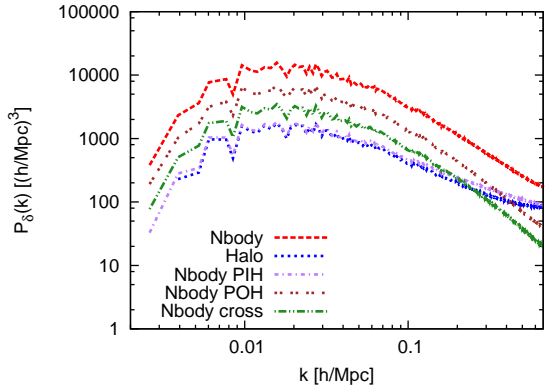}
 \caption{The power spectra for the different components of the N-body simulation compared to the FoF halo matter 
 power spectrum. The red dashed curve shows the matter power spectrum of the N-body simulation evaluated using all of 
 the dark matter particles. The spectra of the particles in the halos (PIH) and the particles outside the halos (POH) 
 are shown with violet dot-dashed and brown dot-dotted curves, respectively. 
 The cross correlation power spectrum between the particles in and outside the halos is represented with a green 
 dashed-dot-dotted curve. The blue short-dashed curve refers to the FoF halo power spectrum.}
 \label{fig:nbody_PH}
\end{figure}

As introduced above, we can divide the particles in the simulation in two different groups: the particles in the 
halos (PIH) and the particles outside the halos (POH). 
In Fig.~\ref{fig:nbody_PH} we show the power spectra for the different contributions (PIH, POH, halos and the 
cross-correlation between the PIH and the POH). 
For the FoF halo identification method described above, most of the particles are outside the halos; the effective 
matter densities are $\Omega_{\rm m,PIH}\approx 0.032$ and $\Omega_{\rm m,POH}\approx 0.218$. 
As a test of consistency, we checked that summing up the three spectra (particles inside and outside the halos and 
its cross-correlation) we obtain the full three-dimensional matter power spectrum.

Both the POH and the PIH show power on all scales, and are strongly correlated with one another. 
While the bias in the power spectrum of the halo particles is stronger, the amplitude shown is considerably lower 
since this reflects the different effective matter densities of each component. 
On linear scales, the spectra of the particles inside and outside the halos agree after a suitable rescaling of the 
amplitude due to the different effective matter density parameters. 
Particles in halos dominate the power seen at high $k$; the slope in the non-linear regime for the PIH spectrum is 
shallower because the non-linear (one-halo) structures contribute to it. 
Not surprisingly, there is very good agreement between the power spectra evaluated from the halo catalogue and the 
particles building the halos. 
The agreement is excellent up to very non-linear scales ($k\approx 0.2$~h/Mpc) where the shot noise of the halos 
becomes important.

\begin{figure}
 \includegraphics[width=0.3\textwidth,angle=-90]{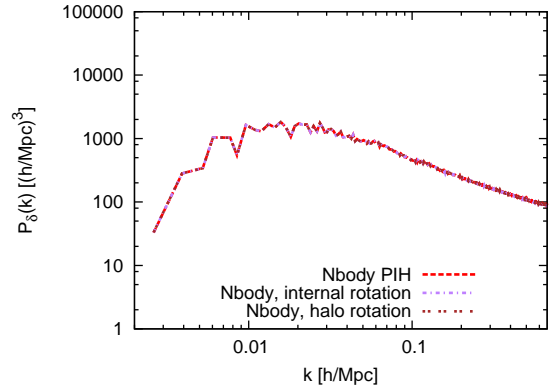}
 \includegraphics[width=0.3\textwidth,angle=-90]{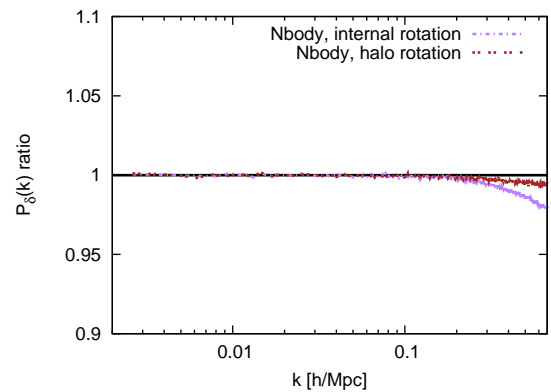}
 \includegraphics[width=0.3\textwidth,angle=-90]{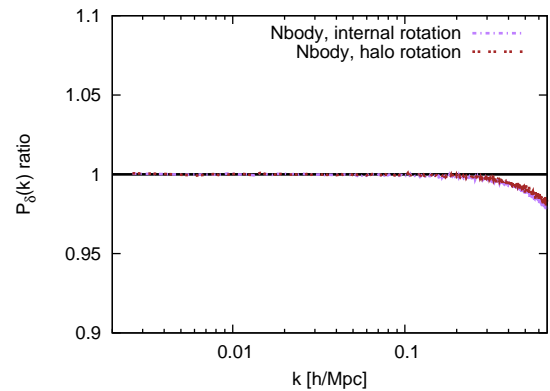}
 \caption{Upper panel: Matter power spectra with rotated halos. The red dashed curve shows the matter power spectrum 
for the N-body simulation using the original particle distribution in the halos. The violet dot-dashed curve 
represents a randomization of the positions of particles inside the halos while the brown dot-dotted curve shows the 
power spectrum when halos are rotated coherently. Middle panel: ratio between the power spectrum after rotations were 
performed and the original one. Bottom panel: ratio between the cross correlation power spectrum between the particles 
in and outside the halos after rotations were performed and the original one.}
 \label{fig:nbody_rotation}
\end{figure}

To further evaluate the influence of the internal structure of halos, we rotate the particles about the halo centres 
as described above, both coherently and incoherently. 
The results are presented in Fig.~\ref{fig:nbody_rotation}, compared to the original PIH spectrum. 
When the halos are rotated coherently, there is very little impact on the power spectrum; a small decrease of power 
at  the non-linear scales (at most 1\%) apparently results from missing small correlations between substructures in 
neighbouring halos. 
When the positions of individual particles are rotated inside the halos, the halo substructures are wiped out and we 
notice a decrement in the matter power spectrum of order 2-3\% for wavelengths $k\approx 0.7$~h/Mpc, the highest 
wavelength we can probe in our analysis.

This shows two things: 
First, the matter power spectrum is almost completely independent of the halo orientation; what matters most are the 
halo masses and their positions with respect to other halos (i.e. their correlation function). 
Second, the power spectrum obtained from the particles in the halos is largely independent of the internal 
substructures, just showing a decrement at the few percent level. 
This shows that most of the power in the internal halo distribution lies in the radial profile rather than in smaller 
substructures. 
The upper and middle panel in Fig.~\ref{fig:nbody_rotation} show that the power in halos is more impacted by the 
incoherent rotations than by the coherent rotations. This demonstrates that significant power is in either the 
ellipticity or the substructure of halos. 
Interestingly also the cross-correlation power spectrum is affected by the rotation of the particles building the 
halos in a similar fashion to the halo power spectrum. Differences are appreciable once again at small scales and 
both rotations have very similar quantitative consequences, albeit in slightly smaller amplitude than the halo power 
spectrum. This is likely due to the induced misalignment between coherent substructures and the cosmic web. It is 
worth reminding the reader that on small scales, the cross-correlation between PIH and POH is considerably smaller 
than the autocorrelation of the PIH itself (see Fig.~\ref{fig:nbody_PH}), so the halo-cosmic web misalignment has a 
relatively small impact on the total power. 
While halo substructures appear to have an effect on the matter power spectrum limited to very small scales, this will 
not be the case any more for the lensing spectrum, as shown in Sect.~\ref{sect:Nbody_Pl}, since this is an integrated 
quantity. In particular, as a consequence, effects of substructures on the lensing power spectrum will be emphasised.

\subsection{Lensing power spectrum}\label{sect:Nbody_Pl}
In this section we will study and compare the lensing signal from the full N-body simulation and from the halo 
catalogue as well.

\subsubsection{Creating the lensing maps} 
In the following we will describe the procedure used to create the lensing maps taking into account the particle 
distribution or the halos in the FoF catalogue. 
Unfortunately, we have the matter distribution at a single redshift only, so we are not able to fully ray-trace 
through an evolving simulation. 
Since the snapshot represents the simulated matter in a volume of 2.4~Gpc/h comoving, we assume that this same box 
spans the matter distribution in the universe between $z=0$ and $z\approx 1$. 
This is obviously not correct because we do not take into account the gravitational evolution of structures, but it 
is still a useful exercise to evaluate the signature of the different components under investigation, in particular 
how substructures will affect the lensing signal.

To create the lensing maps using the particle distribution, we closely follow the procedure outlined in 
\cite{Fosalba2008}, which we refer to for more details. 
There are two major differences between our maps and the convergence maps created by \cite{Fosalba2008}. 
Since we only have one box at a given redshift, we divide the box into slices, and consider each as an independent 
snapshot at a different redshift and sum the contributions of all of them after a suitable weighting due to the 
lensing efficiency. 
The second difference is that we do not cover the whole sky, but just a portion of it; however, since lensing is a 
line of sight effect, this is not an issue.

In the widely used Born approximation \citep{Bartelmann2001,Cooray2002,Refregier2003,Vale2003}, lensing distortions 
are integrated over the unperturbed photon paths and the effective convergence is simply defined as the weighted 
projected surface density:
\begin{equation}\label{eqn:kappa}
 \kappa(\vec{\theta})=\frac{3H_0^2\Omega_{\rm m,0}}{2c^2}\int dr
 \delta(\vec{r},\vec{\theta})\frac{r(r_{\rm s}-r)}{ar_{\rm s}}\;,
\end{equation}
where $a$ is the scale factor, $r$ the radial distance and $r_{\rm s}$ the radial position of the lensing sources. 
In our simulations, we assume all the sources are located at $z_{\rm s}=1$, corresponding roughly to a comoving 
distance of 2.4~Gpc/h. 
Finally, $\vec{\theta}$ represents the angular position on the source and lens plane. Given the effective convergence 
$\kappa(\vec{\theta})$, it is possible to construct shear, deflection angle and lensing potential maps, though here 
we limit ourselves to the study of the effective convergence only.

Since the matter distribution is discrete, we can replace the integral in Eqn.~(\ref{eqn:kappa}) with a summation 
over the lensing planes \citep{Fosalba2008}. The effective convergence at a pixel $(i,j)$ will therefore be:
\begin{equation}
 \kappa(i,j)=\frac{3H_0^2\Omega_{\rm m,0}}{2c^2}\sum_{k}\delta(i,j;k)\frac{r_k(r_{\rm s}-r_{k})}{r_{\rm s}}dr_{k}\;,
\end{equation}
where $dr_{k}$ represents the plane separation (the width of the radial bin in \cite{Fosalba2008}). The density in 
each pixel, for a given lensing plane, $\rho(i,j;k)$, is defined as
\begin{equation}
 \rho(i,j;k)=\frac{m(i,j;k)}{dV(i,j;k)}\;
\end{equation}
where $dV$ corresponds to the volume associated with the pixel and $m(i,j)$ the mass included in the pixel $(i,j)$ 
for each lensing plane $k$. Summing over all the lensing planes, we obtain the final convergence maps.

We checked the numerical stability of our ray-tracing implementation changing the number of lensing planes and the 
number of pixels on each plane, using either eight or sixteen lensing planes and a resolution of 2048$^2$ or 4096$^2$ 
pixels. 
We evaluated the lensing power spectrum for all the realisations and we obtained identical results. 
This confirmed the numerical stability of our implementation. 
In the following we will therefore show results for maps obtained summing up 8 planes and with a resolution of 
2048$^2$ pixels. 
To evaluate the angular scale of our convergence maps, we fix a comoving size on the source plane of 500~Mpc/h. 
This corresponds, in our cosmology, to a field-of-view aperture of $\approx$12\degree, therefore the pixel resolution 
is $\approx 21^{\prime\prime}$. 
We cut a cone through the box with side of 0.5~Gpc/h (comoving) in order to remain in the flat-sky approximation. 
To create more realisations of the effective convergence maps from our single simulation snapshot, we placed the 
observer in six different faces of the box, so that the power spectrum obtained is the mean of six different 
realisations.

So far we have described the algorithm we implemented when the matter distribution is represented by the particles in 
the simulation. 
Something similar can be done to create the lensing maps for the halos of the FoF catalogue. 
The FoF catalogue provides for each halo its mass and location, that we assume to represent the centre of the halo. 
To evaluate the convergence of the halo, we assume that the halos are described by a spherically symmetric NFW 
profile. 
Knowing the mass and the redshift of the halo, we infer the virial radius, the concentration parameter and the scale 
radius and using the equations described in Sect.~\ref{sect:HM}, we evaluate the lensing signal as a function of the 
distance from the centre of the halo. 
Summing up the contribution from all the halos, we obtain the convergence map. 
As above, we verified the robustness of the implementation for the halos varying the number of planes and the map 
resolution; these agree very well, and to be consistent with what we did for the particle distribution, we use eight 
lens planes and a resolution of 2048$^2$ pixels.

We briefly clarify the concentration parameter we adopt for the halos. 
The concentration parameter for the NFW halo model depends on both the mass and the redshift of the halo; for 
consistency, when we evaluated the convergence of the halo fields, we used the appropriate relation discussed in 
Section~\ref{sect:HM}. 
But in our simulation, the halos are all at the same redshift, making therefore the concentration parameter 
effectively dependent only on the mass. 
To check the influence of the concentration parameter on the lensing statistics, we constructed the convergence maps 
in two ways: in the first, we considered the concentration parameter evolving with redshift, while in the second case 
we modify the concentration accordingly with the mass, but assuming the same redshift for all the objects.

\begin{figure}
 \centering
 \includegraphics[width=0.3\textwidth,angle=-90]{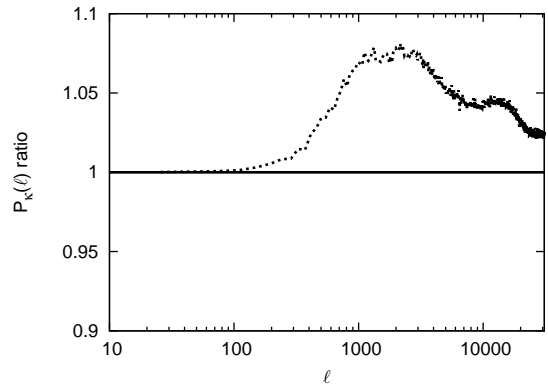}
 \caption{The ratio between the lensing power spectrum with unevolving concentration parameter and the spectrum 
 where the redshift evolution is taken into account.}
 \label{fig:conc_param}
\end{figure}

We show the ratio between the two different implementations in Fig.~\ref{fig:conc_param}. 
On large scales the two spectra are identical, while on small scales they differ by a few percent, with the spectrum 
with the unevolving concentration being slightly higher. 
Larger differences appear at intermediate frequencies, where the spectrum with unevolving concentration parameter is 
roughly 8\% higher than the spectrum where the concentration parameter is a function of mass and redshift. In the 
following we will assume that the concentration varies with mass and redshift and will discuss the impact of this 
assumption.

Another comment is necessary at this point. Comparing our results to better ray-tracing simulations, we need to 
emphasize that there will be additional contributions. Not only the concentration relation will evolve in time, but 
also the mass function will play an important role in changing the number of halos as a function of redshift.

\subsubsection{Comparing the convergence spectra}\label{sect:Kcomparison}

\begin{figure}
 \includegraphics[width=0.3\textwidth,angle=-90]{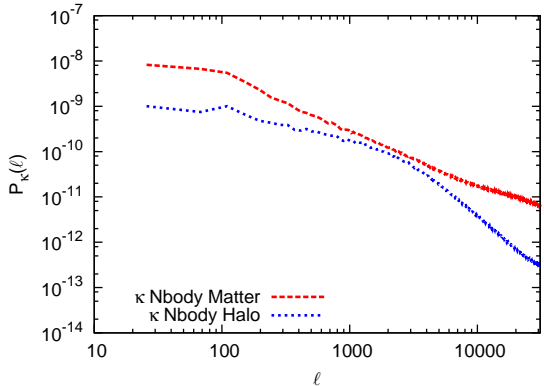}
 \caption{Effective convergence power spectrum for the total matter distribution (red dashed line) and the FoF halo 
 catalogue (blue dotted line).}
 \label{fig:NHpl}
\end{figure}

In Fig.~\ref{fig:NHpl} we compare the N-body lensing power spectrum, obtained using the particles in the simulated 
box, with the spectrum obtained using the halos only. 
The lensing power spectrum is significantly different from the total spectrum; unlike the matter power spectrum, 
they cannot be matched with a simple rescaling to account for the effective matter density. The two spectra have very 
different shapes. 
The lensing spectrum of the halos has a lower amplitude (since $\Omega_{\rm m,h}<\Omega_{\rm m}$) and differs both 
at low and high $\ell$s, while it is comparable around $\ell\approx 2\times 10^{3}$. 
This suggests the importance of the cosmic web - structures not in halos - for the lensing signal, being them the 
primary responsible for the differences of shape and amplitude on large scales for the spectrum obtained using only 
the halos. 

\begin{figure}
 \includegraphics[width=0.3\textwidth,angle=-90]{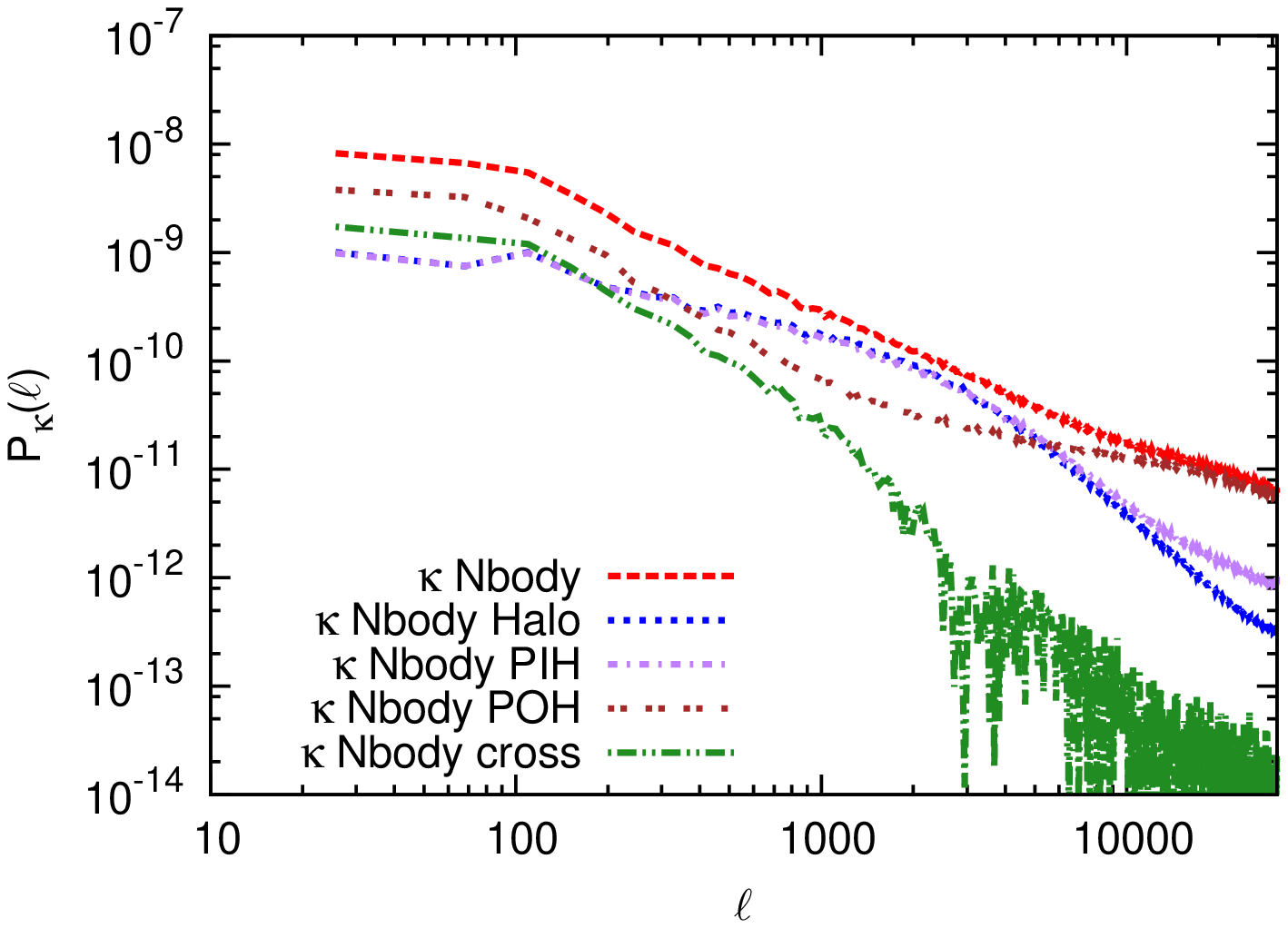}
 \includegraphics[width=0.3\textwidth,angle=-90]{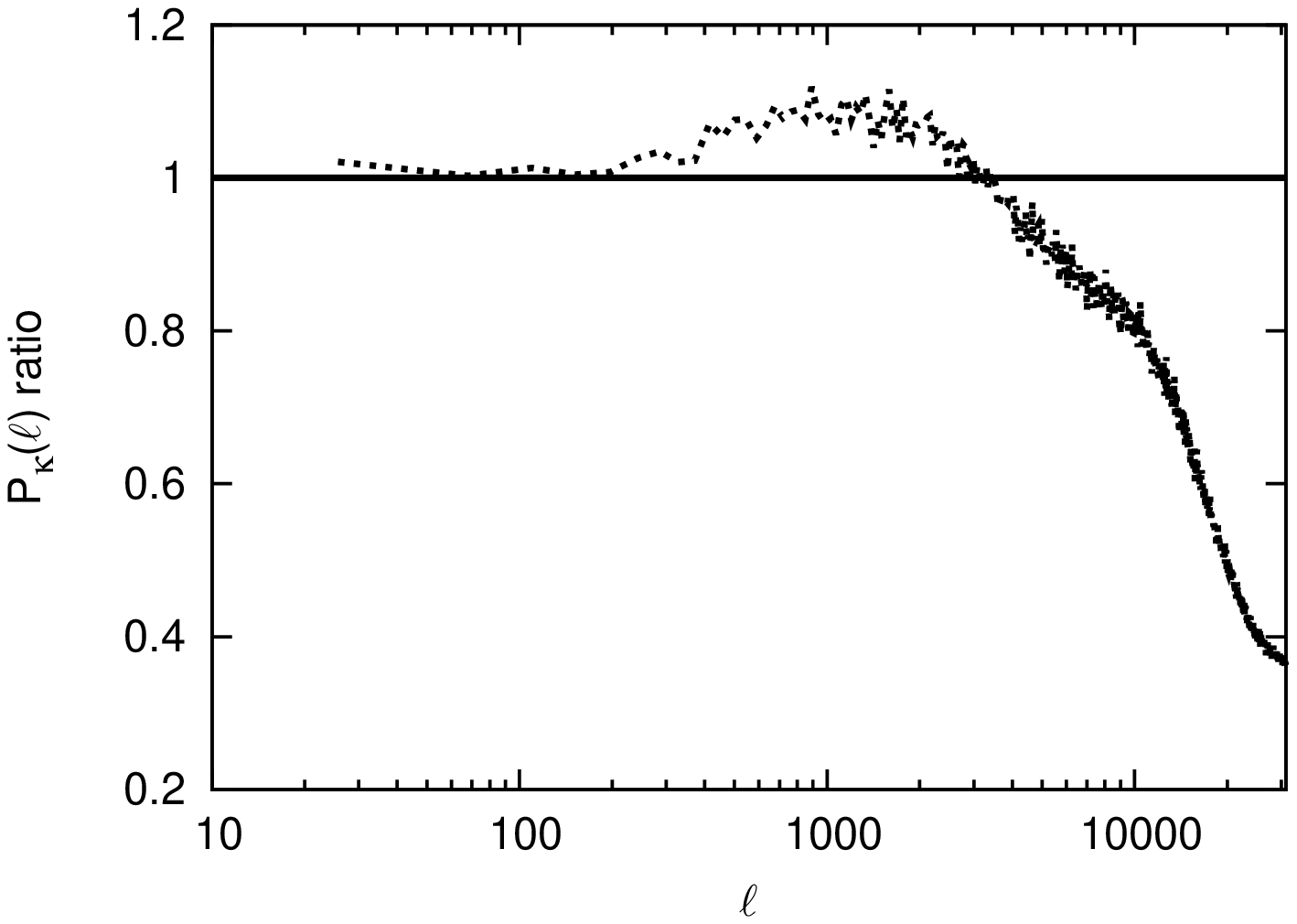}
 \caption{Upper panel: convergence power spectrum for the different components in the N-body simulation and the NFW 
halo catalogue. Lower panel: ratio between the power spectra of the halos and of the PIH. Colours are as in 
Fig.~\ref{fig:nbody_PH}.}
 \label{fig:lnbody_PH}
\end{figure}

\begin{figure}
 \includegraphics[width=0.3\textwidth,angle=-90]{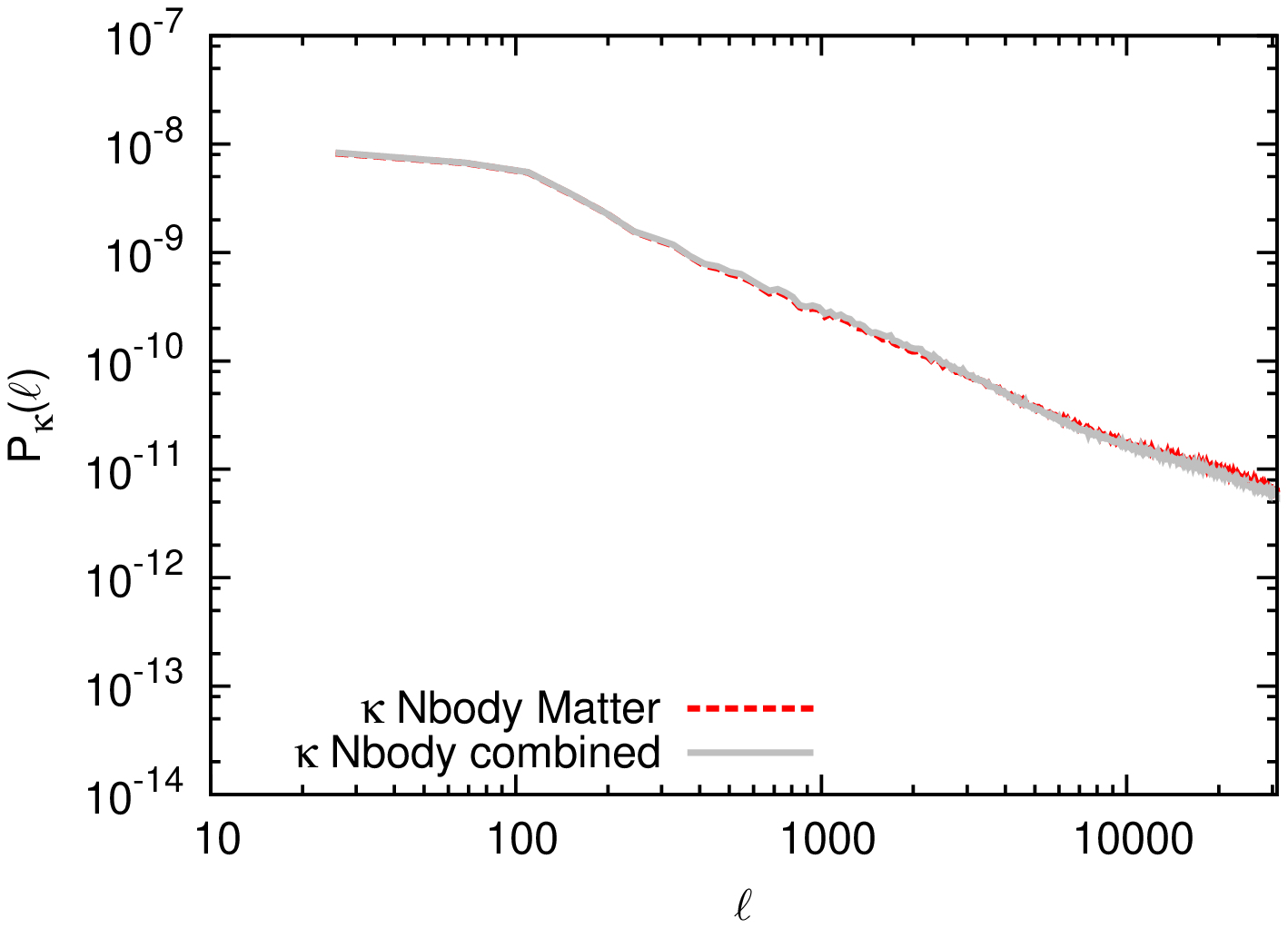}
 \includegraphics[width=0.3\textwidth,angle=-90]{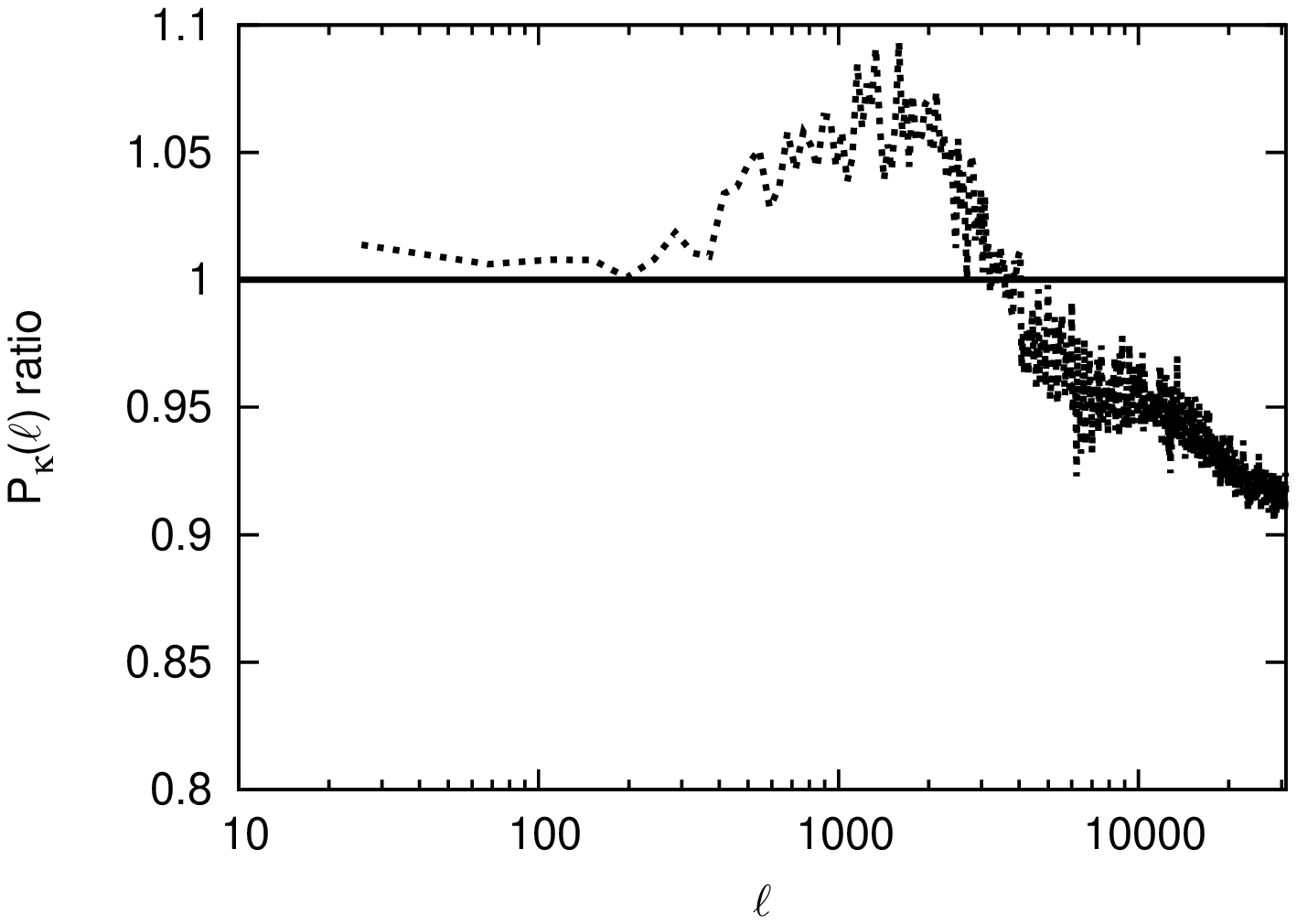}
 \caption{Upper panel: convergence power spectrum from the full particle distribution (red dashed line) and from the 
 combination of the cosmic web (POH) and the NFW halos (grey solid line). Lower panel: the ratio between the spectra.}
 \label{fig:LSS_Halos}
\end{figure}

As above, it is useful to break down the contributions of the particles between those making up the halos (PIH), 
and those particles outside the halos (POH). 
In Fig.~\ref{fig:lnbody_PH}, we show the contribution to the total lensing spectrum of PIH, the POH and of their 
cross-correlation. 
Again, we verified that summing of the spectra of the different components recovers the power spectrum of the full 
field.

The lensing spectrum inferred from the NFW halo catalogue agrees very well with that of the PIH on large scales, up 
to $\ell\approx 200$, but there is an excess of power around $\ell\approx 1000$ of about 10\%. 
However, on small scales we observe a lack of power inferred from the halo catalogue; we find the power dips about 
20\% at $\ell\approx 10^4$, dropping even lower on smaller scales. 
The origin of this complex behaviour is unclear; it must be remembered that the lensing maps are a projection of the 
density maps, and the effects of substructure missing in the halo treatment (as well as associated shot noise) could 
be spread over a range of multipoles.   

Different concentration recipes give very similar results on small scales, but they can impact the excess of power 
seen on intermediate scales. 
The peak seen in the power ratio can change as much as 5\% and shifts towards smaller frequencies when changing the 
concentration relation. When considering FoF halos, the amplitude and the position of the excess is related to the 
assumed concentration relation, but its origin is more complex and related to the asphericity of the FoF halos, as we 
will explain later.

In Fig.~\ref{fig:LSS_Halos} we compare the lensing signal of the full simulation to that combining the NFW halos with 
the matter distribution of particles outside the halos (upper panel), showing their ratio in the lower panel. These 
are largely in agreement, but are unable to reproduce the signal exactly, due to the lack of power on large $\ell$s in 
the NFW halos and the mismatch on intermediate scales (see lower panel in Fig.~\ref{fig:lnbody_PH}). 
Due to the importance of the POH and its cross correlation with the halos, the lack of total power is limited to 10\% 
and changes little over a decade in scale.

We also checked how these results are affected by the definition of a halo itself, by analysing maps created with 
catalogues whose halos had a different particle number threshold. 
All structures having fewer particles were excluded from the halos and the particles building them were absorbed into 
the cosmic web. 
Thresholds of 20, 100, 200, 1000 and 2000 particles, corresponding to a mass threshold of $9\times 10^{11}$, 
$4.6\times 10^{12}$, $9\times 10^{12}$, $4.6\times 10^{13}$ and $9\times 10^{13}~M_{\odot}/h$, were considered. 
The differences between the combined POH plus mock catalogues and the full particle distribution for the halos become 
progressively smaller when fewer halos are used (and the fraction of mass in the POH increases).

\begin{figure}
 \centering
 \includegraphics[width=0.3\textwidth,angle=-90]{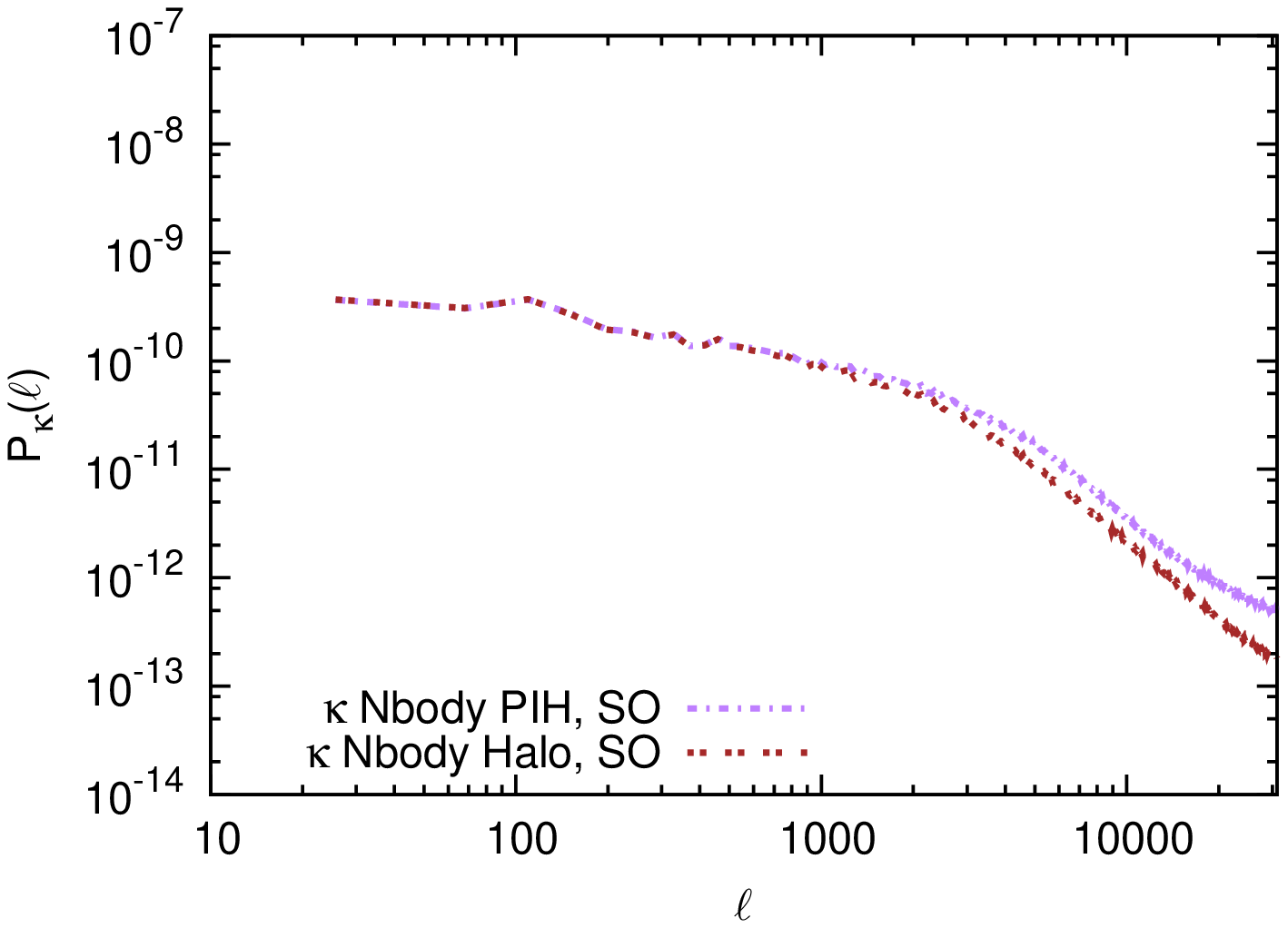}
 \includegraphics[width=0.3\textwidth,angle=-90]{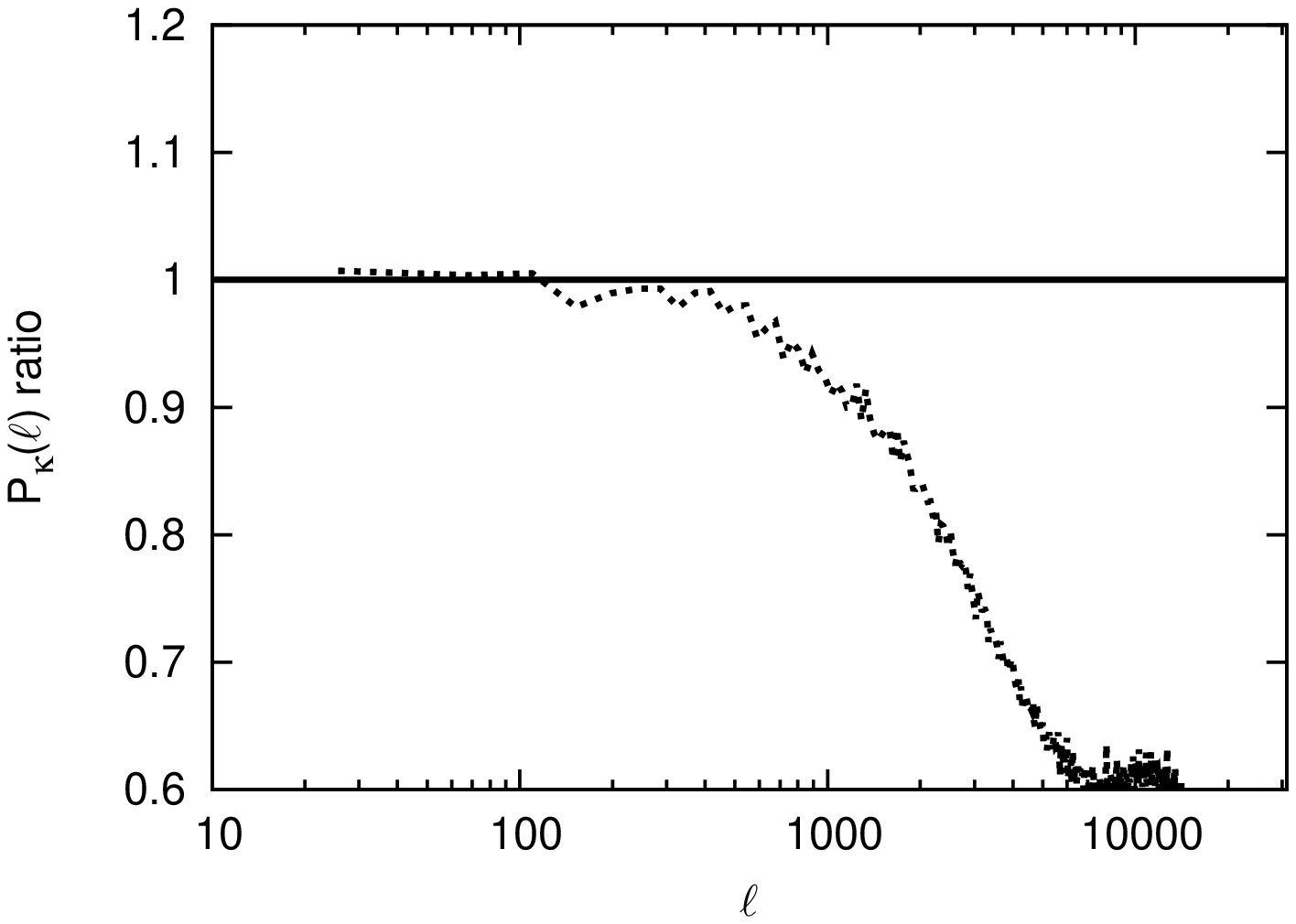}
 \caption{Upper panel: convergence spectrum from the particles building the SO halos and the corresponding NFW ones. 
 Lower panel: ratio between the spectra.}
 \label{fig:SO}
\end{figure}

One might wonder whether the differences using the analytical halos could be related to how the halos were identified 
in the simulations. 
The NFW profile assumes that halos are spherically symmetric objects with a given over-density (in our case 200 times 
the critical density). 
Obviously this is not necessarily true in the FoF halos that they replace. 
To explore whether the spherical symmetry of the halos is an issue, we apply a spherical over-density algorithm on top 
of the FoF halo catalogue to identify the particles in the halos. 
The halos identified in this way are by definition more spherical; we also require the number of particles must be at 
least 20 and the density enclosed in the sphere to be $\rho=200\rho_{\rm c}$. 
These spherical over-density (SO) halos will be smaller (and therefore less massive) and in a lower number with 
respect to the FoF halos. 
To create the SO-halo catalogue from the FoF-halo catalogue, we consider a sphere, centred on the halo centre with 
radius equal to the maximum distance of the particles from the centre of the halo; we then shrink this radius until 
the internal average density is 200 times the critical density. 
If the final number of particles is less than 20 (the minimum number of particles to have an halo) we discard that 
halo and any discarded particles instead contribute to the cosmic web.

In Fig.~\ref{fig:SO} we show the comparison between the lensing signal from the particles building the numerical SO 
halos and the corresponding analytical NFW halos. 
The agreement on large scales is once again achieved, as expected. Up to $\ell \sim 400$ the two spectra are 
practically identical while with the increase of the multipole we notice the usual lack of power, while we find the 
power decreases rather than increases at intermediate scales. 
Comparing Fig.~\ref{fig:SO} with Fig.~\ref{fig:LSS_Halos} we notice that the previously discussed peak disappears. 
This tells us that the FoF halos are responsible for excess of power at intermediate scales. When we compare a 
spherical halo with a FoF halo, we observe that the latter will have particles outside the modelled spherical area, 
creating overdense regions that by construction do not exist in the case of the SO halos. These regions are 
responsible for the excess of power at intermediate scales.

\subsubsection{Rotating the halo particles}\label{sect:subsPl}

We next explore the impact of rotating halo particles, either randomly or coherently, on the lensing convergence 
spectrum. 
Recall that for the power spectrum, these rotations had relatively little impact; however, by suppressing halo 
substructures, the incoherent rotations did suppress the large $k$ spectrum (Fig.~\ref{fig:nbody_rotation}).

\begin{figure}
 \centering
 \includegraphics[width=0.3\textwidth,angle=-90]{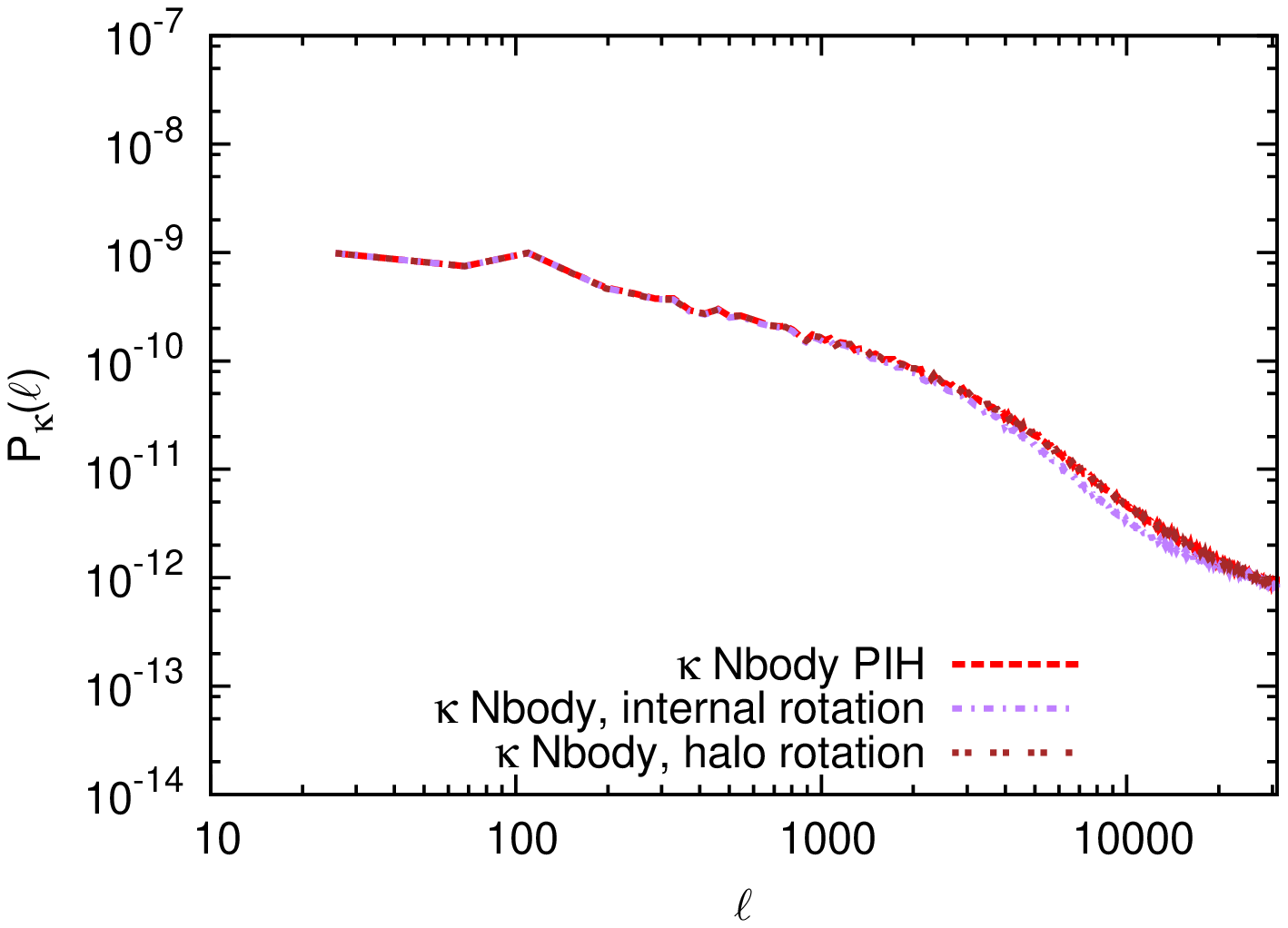}
 \includegraphics[width=0.3\textwidth,angle=-90]{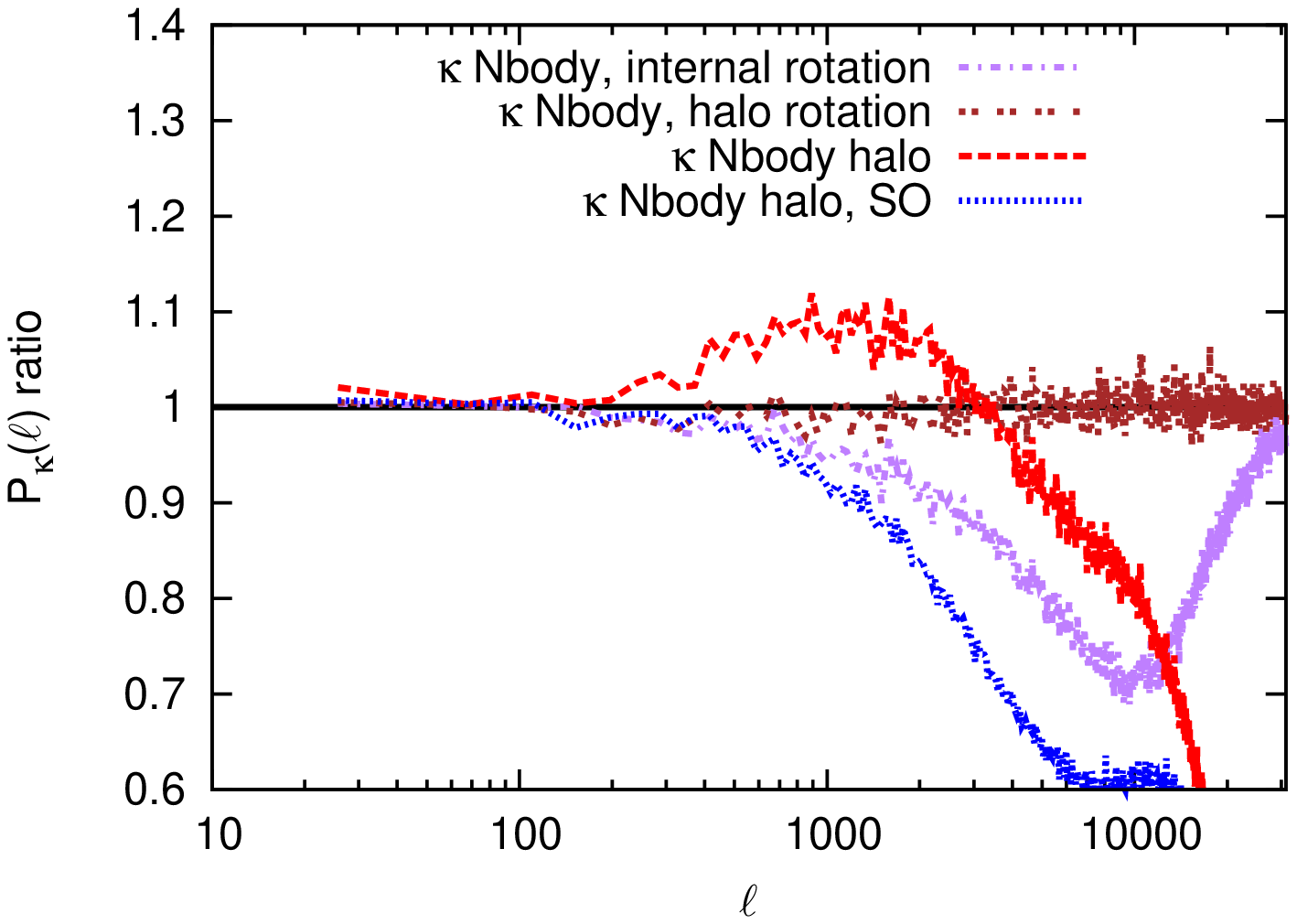}
 \caption{Upper panel: Comparison between the effective convergence power spectra for the original simulation 
 (red dashed curve) limited to the particles in the halos, and the same particle distribution but with a rotation 
 about the centre of each halo (violet dot-dashed curve) or with randomization of the particles' positions (brown 
 dot-dotted line). Lower panel: ratio between the different rotated matter distribution spectra and the original one. 
 The red dashed and blue dot-dotted lines represent the spectra of the NFW halos, as a ratio to the FoF or SO 
 simulated halos, respectively.}
 \label{fig:kappa_rot}
\end{figure}

The impact of the rotations for the lensing spectrum is shown in Fig.~\ref{fig:kappa_rot}. 
There is no appreciable difference on large scales; both kinds of rotations remain in very good agreement with the 
original PIH spectrum. 
However, on small angular scales, where one-halo effects become important (above $\ell\approx 1000$), there is a 
notable lack of power for the incoherent rotations that becomes as large as $\approx 30\%$ before the shot noise 
contribution becomes dominant at $\ell\approx 10,000$.  This indicates that the halo substructures play a 
larger role in the determination of the lensing maps than they do for the density maps, due to closer substructures 
being projected into larger angular scales. 
It may be possible to improve the halo model results by adding substructure to the assumed profiles to reflect the 
mass and profile distributions \citep{Giocoli2008,Giocoli2010} of the satellite sub-halos. Alternatively, one could 
create a library of halos including substructures and assign them randomly to the mock halos.

It is interesting to compare these results to the case where the particles in a halo are rotated coherently, 
preserving the halo substructures. 
In this case (also shown in Fig.~\ref{fig:kappa_rot}), the power is largely preserved compared to the original 
simulations (as also seen in the matter power spectrum); this indicates that it is indeed the halo substructure, 
rather than the halo profile, that makes the differences seen for the incoherent rotations. 
Mixed in with the effect of substructure will also be the ellipsoidal distortions seen in realistic halos; while 
spherically symmetric objects have no preferred direction, the effective convergence of the halos (ellipsoidal rather 
than spherical) will change significantly if the projection is made along the longest or shortest axis. Thus the 
ellipsoidal structure of halos could add additional variance in the lensing power spectrum.

From the analysis of Fig.~\ref{fig:kappa_rot} we can draw another important conclusion. By comparing the curves 
showing the ratio between the lensing power spectrum based on the FoF halo catalogue (red dashed curve) and on the SO 
halo catalogue (blue dot-dotted curve), we notice that the peak at $\ell\approx 1000$ disappears, clearly showing 
that its presence can be related to the non-spherical halos typical of FoF catalogues as well as bridged halos 
(unbounded particles included by the FoF algorithm). We want to stress out that the peak at $\ell\approx 1000$ is the 
effect of using FoF halo catalogues instead of SO halo catalogues on lensing maps.

\begin{figure}
 \includegraphics[width=0.3\textwidth,angle=-90]{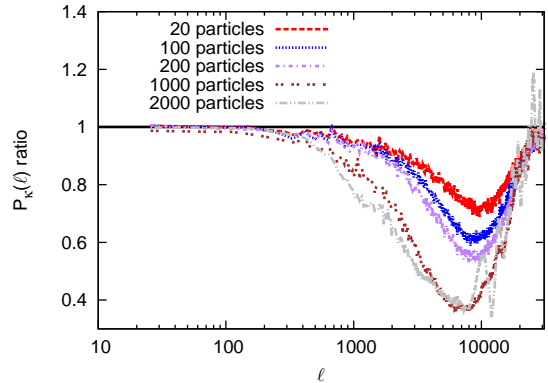}
 \caption{Ratio between the power spectra of the halos with internal substructures washed out and the original 
 convergence power spectrum of the numerical halos for different particle number (mass) threshold. Different colour 
 and line styles correspond to different particle thresholds. The red dashed line corresponds to halos with at least 
 20 particles; the blue dotted (violet dot-dashed) line to halos with at least 100 (200) particles; the brown 
 dot-dotted (grey dashed dot-dotted) line to halos with at least 1000 (2000) particles. A threshold of 20, 100, 200, 
 1000 and 2000 particles corresponding to a mass threshold of $9\times 10^{11}$, $4.6\times 10^{12}$, 
 $9\times 10^{12}$, $4.6\times 10^{13}$ and $9\times 10^{13}~M_{\odot}/h$, respectively.}
 \label{fig:kappa_rot_p}
\end{figure}

Finally, we investigate how the substructure contribution depends on the masses of halos we consider. 
In Fig.~\ref{fig:kappa_rot_p}, we investigate how the effect of substructures is affected by the different definition 
of minimum halo mass considered, plotting the ratio of the effective convergence power spectrum of halos whose 
internal substructures are washed out to the PIH power spectrum for different halo mass thresholds. 
Massive halos are rarer but are rich in substructures and contribute more to the total halo mass. Thus, it is not 
surprising that erasing substructures in massive halos leads to a bigger suppression than when all the halos are 
included. 
Changing the threshold by a factor of fifty doubles the fraction of suppressed power and shifts it to larger angular 
scales. 
Given the rarity of very massive objects, we do not see appreciable differences with a threshold of 1000 or 2000 
particles.

Our inability to reproduce the full simulation results should be compared with the findings of \cite{Petkova2013}.  
In that work, the authors present an extension of the raytracing code GLAMER \citep{Metcalf2013} to the case of 
multiple deflection planes and compare their findings with several existing works. 
In particular they assemble spherically symmetric NFW halos drawn from the Sheth \& Tormen mass function and compare 
the probability distribution functions for deflection angle, shear and effective convergence to N-body results from 
\cite{Pace2011}, finding good agreement. 
However, in \cite{Pace2015} it was shown that for high values of the effective convergence, the raytracing 
simulations underestimated analytical or semi-analytical solutions. 
Thus the agreement may have been due to the suppressed power observed here for NFW halos compensating for the loss of 
resolution in the procedure adopted to build the lensing planes 
\citep[for a more complete discussion see][]{Pace2015}. 
Note that, because we directly use the particle distribution to evaluate the effective convergence rather than 
evaluating it as the Laplacian of the 
lensing potential, we expect the resolution issues here to be less severe than in \cite{Pace2007.1}, \cite{Pace2011} 
and \cite{Pace2015}.

To conclude, our results show that both the cosmic web and the halo substructures are necessary to fully reproduce 
the lensing statistics, suggesting that halo based approaches may be seriously flawed for lensing studies.

\section{2LPT methods}\label{sect:2LPT}
In this section we focus on the comparison between N-body simulations and simulation and halo catalogues based on the 
Lagrangian Perturbation Theory (LPT).

\subsection{Lagrangian perturbation theory}\label{sect:2LPT_theory}
Lagrangian perturbation theory (LPT) is a very useful mathematical approach used to describe perturbations in the 
matter density field and to create initial conditions for cosmological N-body simulations 
\citep{Buchert1989,Moutarde1991,Bouchet1992,Buchert1992,Buchert1993,Buchert1994,Hivon1995,Catelan1995d}. 
In the LPT formalism, the current Eulerian position $\vec{x}$ of a given mass element is related to its initial 
Lagrangian position $\vec{q}$ via a displacement vector field $\vec{\Psi}(\vec{q})$ according to the relation
\begin{equation}\label{eqn:disp}
 \vec{x}=\vec{q}+\vec{\Psi}(\vec{q})\;.
\end{equation}

Similarly to what it is usually done for the Eulerian perturbation theory (EPT), one can make a power series 
expansion of the displacement field $\vec{\Psi}$ as a function of the linear over-density field $\delta_{\rm L}$:
\begin{equation}\label{eqn:Psi}
 \vec{\Psi}=\vec{\Psi}^{(1)}+\vec{\Psi}^{(2)}+\vec{\Psi}^{(3)}+\vec{\Psi}^{(4)}+\ldots\;,
\end{equation}
and $\vec{\Psi}^{(n)}$ is related to the $n$th order expansion of the linear matter density field. 
LPT up to third order has been discussed in several works in literature 
\citep{Buchert1993,Buchert1994,Hivon1995,Catelan1995d} and recently calculations at the fourth order were carried out 
by \cite{Tatekawa2013}. Here, we will limit ourselves to the second-order treatment. 
Translating from Eulerian to Lagrangian coordinates, we can obtain the LPT equations at each order by substituting 
Eqn.~(\ref{eqn:Psi}) into Eqn.~(\ref{eqn:disp}) and the equation of motion for particles
\begin{equation}
 \frac{d^2\vec{x}}{dt^2}+2H\frac{d\vec{x}}{dt}=-\frac{1}{a^2}\vec{\nabla}\phi\;,
\end{equation}
where $a$ is the scale factor, $H$ the Hubble expansion rate and $\phi$ the gravitational potential. 

At first order we obtain the well-known Zel'dovich approximation \citep{ZelDovich1970}:
\begin{equation}\label{eqn:1LPT}
 \vec{\nabla}_{\vec{q}}\cdot\vec{\Psi}^{(1)}=-D_{1}(a)\delta(\vec{q})\;.
\end{equation}
In Eqn.~(\ref{eqn:1LPT}), $D_{1}(a)$ represents the linear growth factor and $\delta(\vec{q})$ is the Gaussian 
density field imposed by the initial conditions.
Second-order LPT (2LPT) corrects the Zel'dovich approximation by taking into account gravitational tidal forces and 
the second order displacement $\Psi^{(2)}$ to find
\begin{equation}
 \vec{\nabla}_{\vec{q}}\cdot\Psi^{(2)}=\frac{1}{2}D_{2}(a)\sum_{i\neq j}
 \left(\Psi_{i,i}\Psi_{j,j}-\Psi_{i,j}\Psi_{j,i}\right)\;,
\end{equation}
where $D_{2}$ is the second order growth factor and for a wide range of $\Omega_{\rm}$ is proportional to $D_{1}^{2}$ 
\citep{Bouchet1995}.

At first and second order, it is convenient to define two different gravitational potentials $\phi^{(1)}$ and 
$\phi^{(2)}$ since the solutions are curl-free. 
The displacement fields are related to the gravitational potentials via the relation
\begin{equation}\label{eqn:psi_phi}
 \vec{\Psi}^{(n)}=\vec{\nabla}_{\vec{q}}\phi^{(n)}\;,
\end{equation}
with $n=1,2$. 
Using the relations in Eqn.~\ref{eqn:psi_phi}, the solution up to second order is
\begin{eqnarray}
 \vec{x}(\vec{q}) & = & \vec{q}-D_{1}\vec{\nabla}_{\vec{q}}\phi^{(1)}+D_{2}\vec{\nabla}_{\vec{q}}\phi^{(2)}\\
 \vec{v} & = & -D_{1}f_1\vec{\nabla}_{\vec{q}}\phi^{(1)}+D_{2}f_2\vec{\nabla}_{\vec{q}}\phi^{(2)}\;,
\end{eqnarray}
where $\vec{v}=\frac{d\vec{x}}{dt}$ is the peculiar velocity and $t$ the cosmic time. Finally, $f_{\rm i}$ is the 
logarithmic derivative of the growth factor, $f_{i}=\frac{d\ln D_{i}}{d\ln a}$.

\subsection{2LPT halo catalogues}\label{sect:2LPT_sim}
With the formalism outlined in Sect.~\ref{sect:2LPT_theory}, it is possible to generate the 2LPT displacement and 
evolve particles according to it. From these, halos can be identified and rescaled to provide fast halo catalogues 
(PTHalos); we briefly describe these here and refer to \cite{Manera2013,Chuang2015,Manera2015} for more details.

The algorithm used to create the second order displacement $\vec{\Psi}^{(2)}$ is described in \cite{Scoccimarro1998}; 
it takes advantage of the fast Fourier transform (FFT) technique. 
\cite{Coles1993,Scoccimarro2002} used a cut-off in the linear matter power spectrum and found the halos using merger 
trees. 
For these simulations halos are identified with a standard friends-of-friends (FoF) technique \citep{Davis1985} with 
a modified linking length. 
The theoretical derivation for the evaluation of the modified value for the 2LPT linking length is based on the 
spherical collapse model for the 2LPT and we refer to \cite{Manera2013} for an exhaustive explanation. 
The linking length used for the 2LPT simulations is calibrated on the LasDamas N-body simulations and its value is 
$b=0.38$, significantly larger than the standard $b=0.2$ commonly used in N-body simulations.

The FoF algorithm identifies halo positions and masses in the 2LPT realisations. 
However, the 2LPT masses need to be calibrated by abundance matching to match a fiducial mass function. 
In particular, the \cite{Tinker2008} mass function was used to calibrate the 2LPT halo masses.

As for the LasDamas simulations used previously, 2LPT halos are identified at $z=0.52$. 
The cosmology adopted is, as before, a flat $\Lambda$CDM model with $\Omega_{\rm}=0.25$, $\Omega_{\Lambda}=0.75$, 
$\Omega_{\rm b}=0.04$, $h=0.7$, power spectrum normalization $\sigma_8=0.8$ and spectral index $n_{\rm s}=1$. The  
simulated box is 2.4~Gpc/h comoving on a side with $1280^3$ dark matter particles. In the following we will consider 
objects with at least 20 particles.

\begin{figure}
 \centering
 \includegraphics[width=0.3\textwidth,angle=-90]{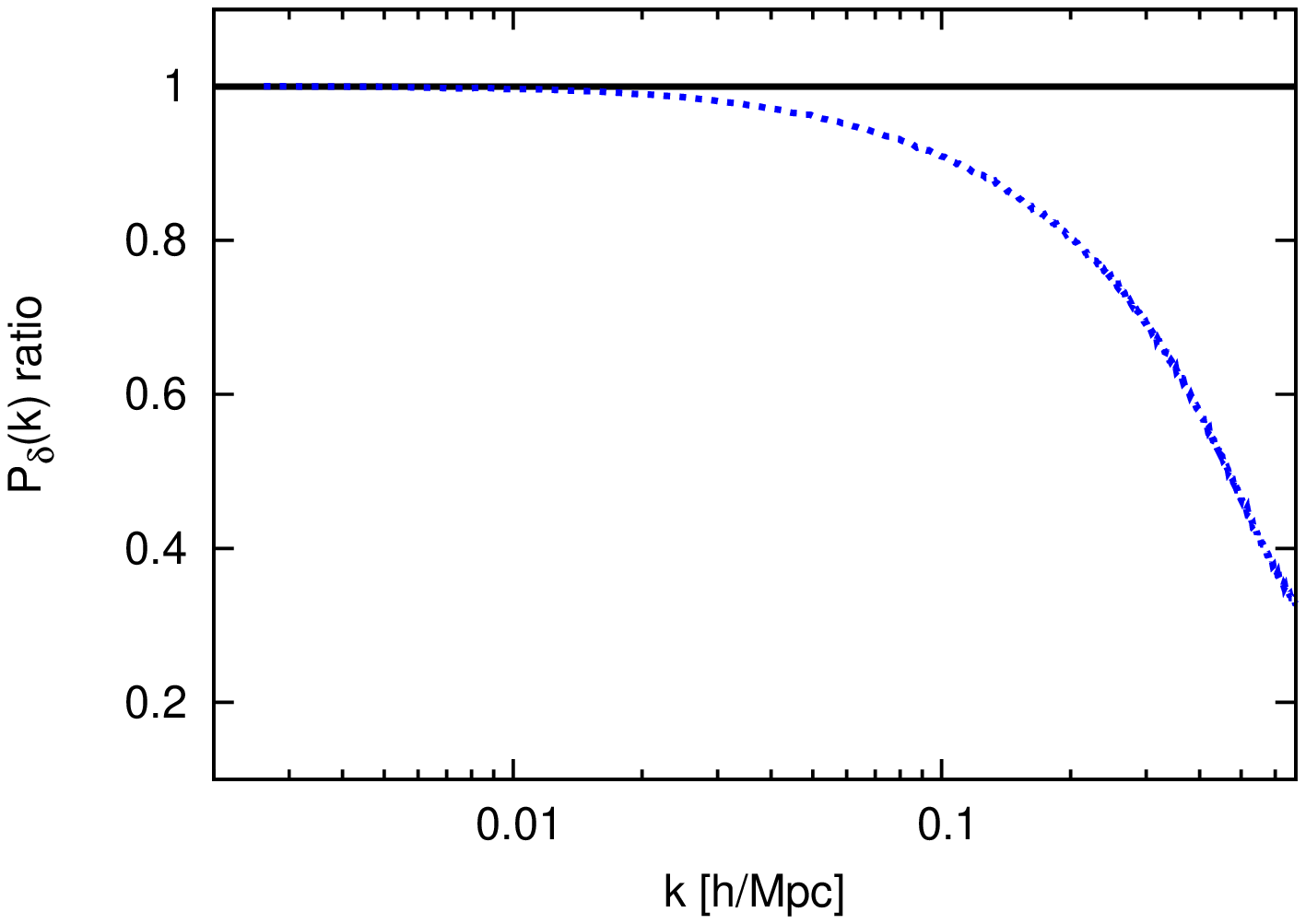}
 \includegraphics[width=0.3\textwidth,angle=-90]{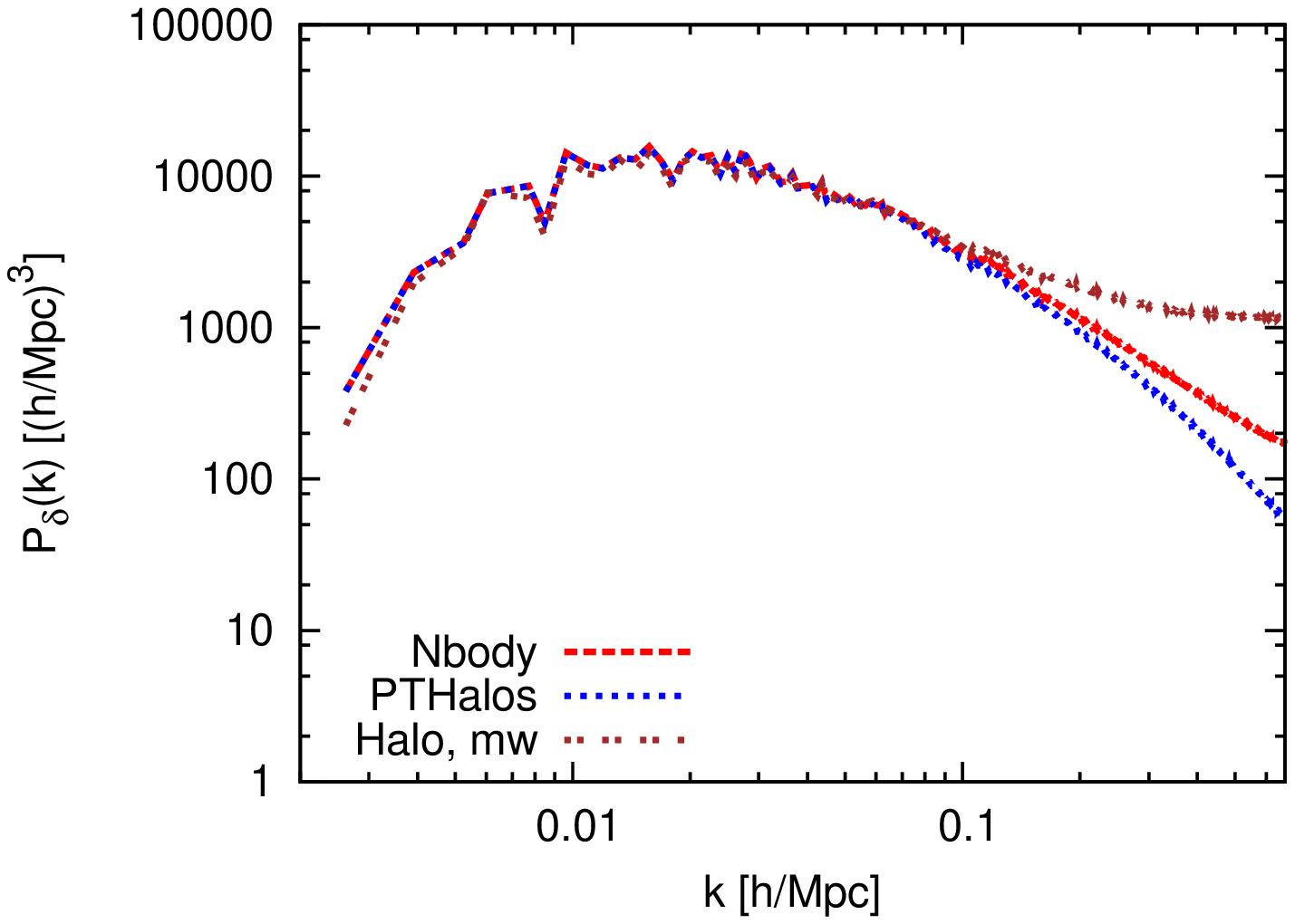}
 \caption[2LPT spectrum]{Upper panel: Ratio between the 2LPT and the N-body matter power spectra (blue short-dashed 
 curve). Lower panel: Comparison of the matter and halo power spectrum for the 2LPT simulation. The red dashed and 
 blue short-dashed curves represent the N-body and the 2LPT matter power spectra, respectively, while the brown 
 dot-dotted curve represents the halo power spectrum with mass weight.}
 \label{fig:2ps}
\end{figure}

\subsection{3D matter power spectrum}\label{sect:2LPT_Pk}
We first study the matter power spectrum following the same procedure outlined in Sect.~\ref{sect:Nbody_Pk}. 
As for the full simulations, we use the NGP \citep{Hockney1988} algorithm to assign dark matter particles to the grid 
and we use 512 points for the Fourier transform.

In the upper panel of Fig.~\ref{fig:2ps} we first compare the full N-body spectrum to that inferred from the 2LPT 
particles, showing their ratio. As expected, the 2LPT spectrum agrees very well with the N-body spectrum on large 
scales, but the spectra differ considerably on small scales where fully non-linear effects are absent in the 2LPT 
treatment. The power deficit is considerable on the smallest modes of the simulation, dropping to 20\% of the full 
simulation. As we will see in the next section (see Sect.~\ref{sect:2LPT_Pl}), this inaccuracy will have tremendous 
effects on the lensing power spectrum.

It is necessary to find alternative methods to enhance the small scale power in the 2LPT simulations. 
In the halo method, the underdeveloped 2LPT halos are replaced by more realistic halo masses. 
As pointed out by \cite{Manera2013}, to correctly reproduce the fiducial mass function, it is necessary to calibrate 
the mass of the FoF halos. 
We therefore compare the halo power spectrum with the total matter power spectrum in the bottom panel of 
Fig.~\ref{fig:2ps} where we refer to the caption for the different line styles and colours. 
As done for the FoF halos in the N-body simulation, we evaluate the halo power spectrum weighting the halos either 
by their (corrected) mass, or simply giving all equal weight.

The halo power spectra reproduce, to large extent, the matter power spectrum on large scales. 
We checked that rescaling the halo power spectrum by an appropriate factor, well reproduces the expected matter power 
spectrum. 
The halo power spectrum with mass weighting is indeed very close to the total one due to a rescaling in the matter 
density parameter.

\begin{figure}
 \centering
 \includegraphics[width=0.3\textwidth,angle=-90]{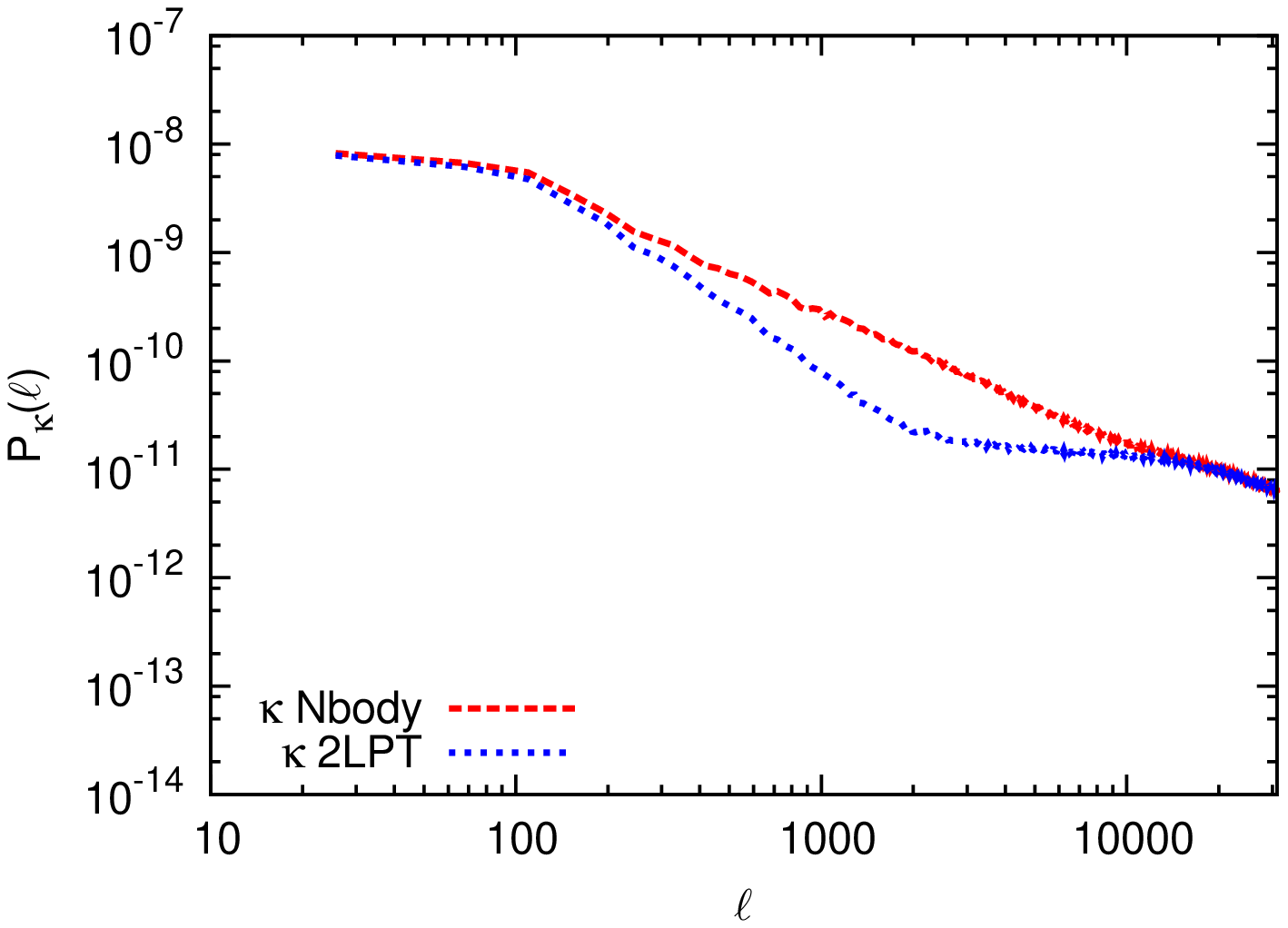}
 \includegraphics[width=0.3\textwidth,angle=-90]{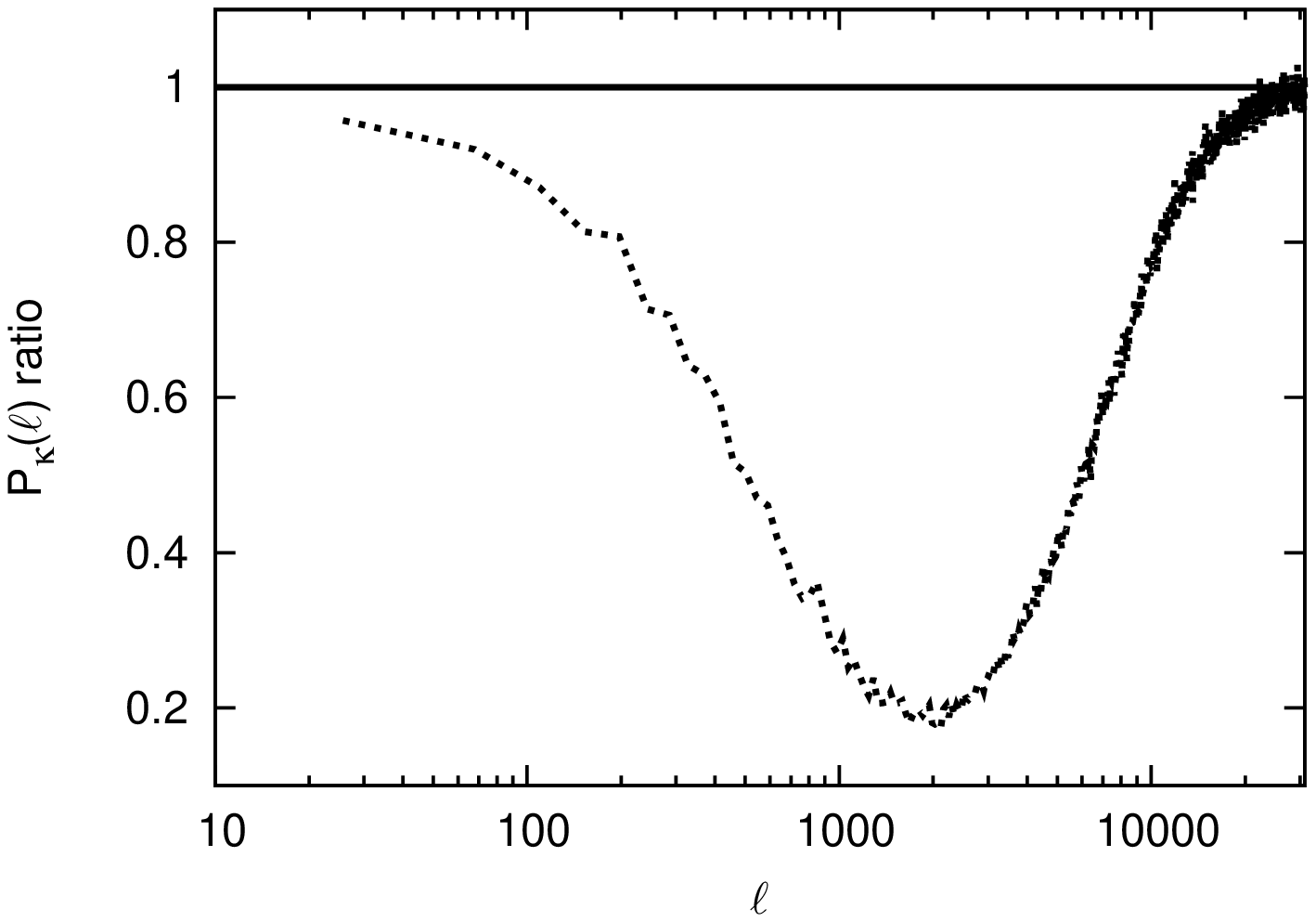}
 \caption{Upper panel: Effective convergence for the N-Body simulation (red dashed curve) and the 2LPT simulation 
 (blue short-dashed curve). Solid line shows the shot-noise power spectrum. Lower panel: ratio between the 
 particle-based 2LPT simulation and the N-body simulation.}
 \label{fig:2pl}
\end{figure}

\subsection{Lensing power spectrum}\label{sect:2LPT_Pl}
We next analyse the lensing power spectrum in the 2LPT simulations, again beginning with the 2LPT particle 
distribution and then seeing how things change using FoF halos derived from the simulations.

In Fig.~\ref{fig:2pl} we show the lensing power spectrum using all the particles in the simulations for sources at 
$z_{\rm s}=1$ for both the N-body and the 2LPT simulation. 
The missing power seen in the matter power spectrum (due to the lack of non-linear evolution in the 2LPT simulations) 
is even more prevalent in the effective convergence power spectrum. 
For very low $\ell$ the agreement is still very good, but the power deficit is significant on intermediate scales, 
beginning as low as $\ell\approx 100$. By $\ell\approx 1000$, the power drops to 20\% of the full simulation; this 
is as bad as the deficit seen in the density spectrum only at the highest $k$ modes. 
Beyond $\ell\approx 1000$, the lensing spectrum flattens out at the expected shot nose level. 
It is perhaps not surprising that the missing non-linear power of the 2LPT simulations is disastrous for predictions 
of the lensing statistics.

\begin{figure}
 \centering
 \includegraphics[width=0.3\textwidth,angle=-90]{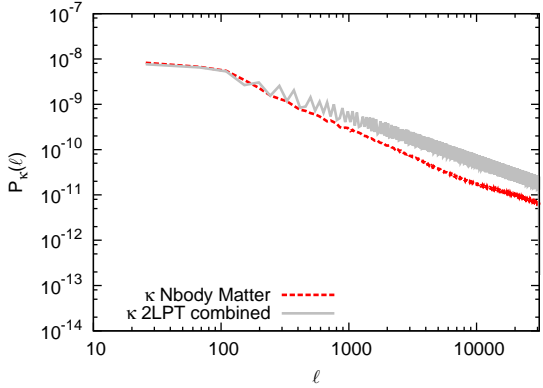}
 \caption{Convergence power spectrum from the N-body full particle distribution (red dashed line) and from the 
 combination of particles building the cosmic web and spherical symmetric halos (grey solid line) of the 2LPT 
 simulation.}
 \label{fig:2lpt_combined}
\end{figure}

Better predictions might be derived using the associated NFW halo catalogues, where the halo masses have been 
rescaled to match the mass distribution observed in full N-body simulations; boosting the halo masses should make 
the deficit shown in Fig.~\ref{fig:2pl} less severe. 
In Fig.~\ref{fig:2lpt_combined}, we show the spectrum of the PTHalos plus the POH in the cosmic web. 
We indeed see that the power decrement of the 2LPT particles has disappeared, but the mass rescaling makes the signal 
too high over a range of frequencies. 
While on large scales the agreement is excellent, on scales with $\ell\gtrsim 100$ we instead note an excess of power 
with respect to the total power measured from the N-body simulation. 
This is due to the mass correction we applied to the halos before evaluating their lensing signal, so that the mass 
of the rescaled halos added is greater than the PIH initially removed. This shows that PTHalos are clustered 
differently.

\begin{figure}
 \centering
 \includegraphics[width=0.3\textwidth,angle=-90]{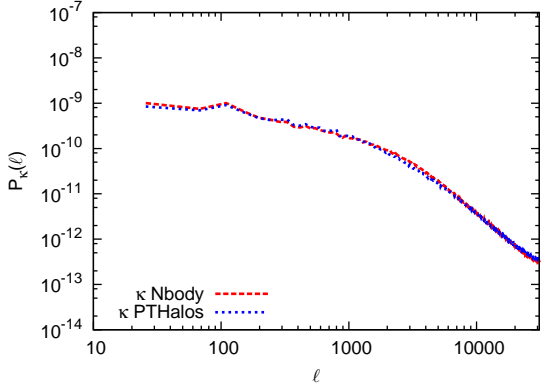}
 \caption{Effective convergence for the N-body halos (red dashed curve) and the 2LPT PTHalos (blue short-dashed 
 curve). Halos are described by a spherical symmetric NFW density profile.}
 \label{fig:2lpt_nbody_halo}
\end{figure}

In Fig.~\ref{fig:2lpt_nbody_halo} we compare the effective convergence power spectrum of the halo catalogues for the 
N-body and the 2LPT simulations, taking into account the mass correction for the 2LPT halos. 
While the mass correction is able to populate the halos with galaxies following the HOD prescription 
\citep{Manera2013}, the effective convergence for halos from the 2LPT catalogue is initially lower than the spectrum 
of 
the halos from the N-body catalogue on all scales.  
However, the total mass in the N-body halos is still greater ($\approx 30\%$) than that in the 2LPT halos even after 
the masses are individually rescaled.
When the effective convergence power spectrum is rescaled by the matter density parameter squared, there is good 
agreement 
with the N-body lensing spectrum, as shown in the figure.  
However, the spectra do not match over the whole range of frequencies for the halo lensing power spectrum; there 
remains a difference in shape, with the 
N-body having more power at $\ell \sim 2000$.   
This might reflect the fact that the mass remapping is not perfect, leading to a somewhat different mass weighting of 
the clustering of the PTHalos than in the N-body simulation.

\begin{figure}
 \centering
 \includegraphics[width=0.3\textwidth,angle=-90]{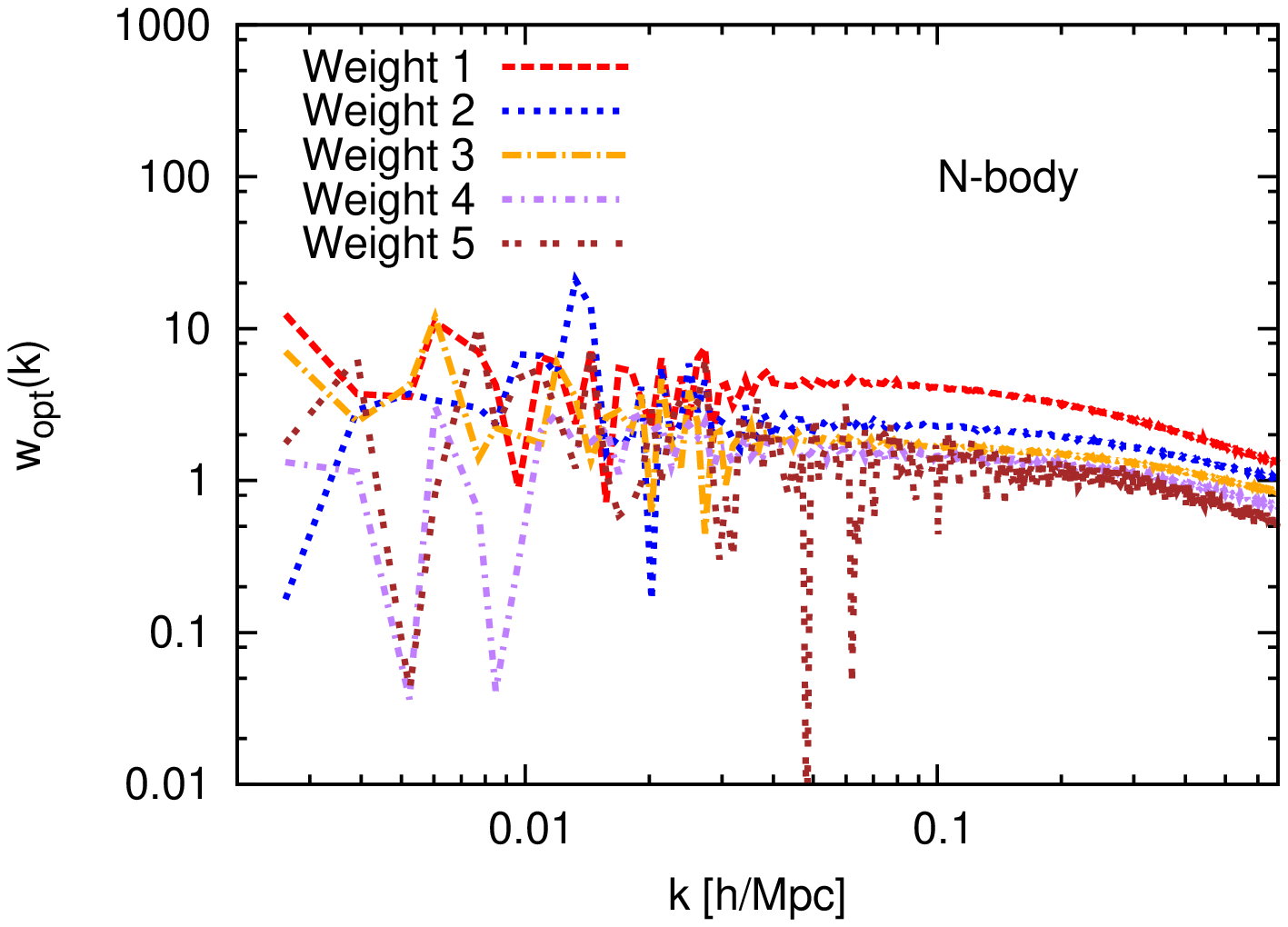}
 \includegraphics[width=0.3\textwidth,angle=-90]{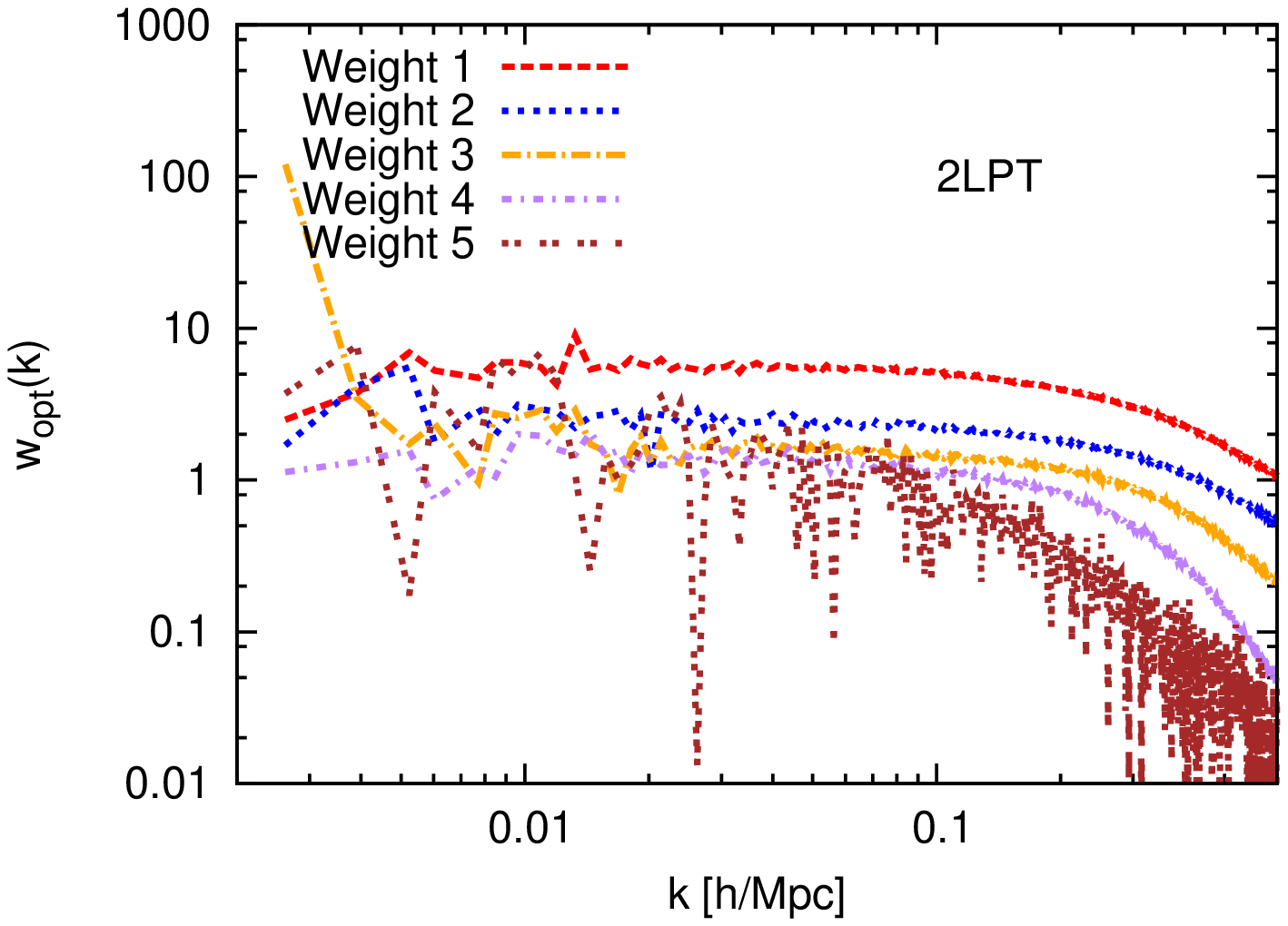}
 \caption{Optimal weight for the five different mass bins as a function of the wavelength $k$. From top to bottom we 
 show the optimal weight for bin 1 (red dashed line), 2 (blue dotted line), 3 (orange dot-dashed curve), 4 (purple 
 dot-short-dashed curve) and 5 (brown dot-dotted curve). Upper (lower) panel shows the weights for the N-body (2LPT) 
 simulations.}
 \label{fig:weight}
\end{figure}

\section{Inclusion of a bias weight}\label{sect:CBS}
Given the limitations of simply using the mock-catalogues when trying to reproduce the results from the N-body 
simulations, in particular those for weak lensing, we next investigate whether these issues can be alleviated by 
using an optimal weight to reconstruct the underlying dark matter distribution from the halo catalogue. 
To do so, we closely follow the work by \cite{Cai2011} and we refer to them for further details. 
The idea is to see whether lensing studies can benefit from a mass-dependent weighting used to reconstruct the three 
dimensional matter field starting from a mock catalogue.

The halo distribution represents a biased sample of the underlying total matter distribution $\delta_{\rm m}$ and the 
goal is to reconstruct the density only using the information we can extract from the mock catalogue. 
We divide our mock catalogue into mass bins and indicate with $\delta_{i}$ the fluctuation in each bin. 
The estimator $\hat{\delta}_{\rm m}$ for the matter density field is then
\begin{equation}\label{eqn:CBSdelta}
 \hat{\delta}_{\rm m}=\sum_{i}w_{i}\hat{\delta}_{i}\;,
\end{equation}
where $w_{i}$ represents the weight we use for each mass bin to reconstruct the dark matter field starting from the 
mock catalogue. 
The power spectrum for the mass fluctuation $P$, the bias vector $\boldsymbol{b}$ and the covariance matrix 
$\boldsymbol{C}$ are defined as
\begin{eqnarray}
 \langle\delta_{\rm m}^2\rangle & = & P\;,\\
 \langle\delta_{\rm m}\delta_{i}\rangle & = & b_iP\;,\\
 \langle\delta_{i}\delta_{j}\rangle & = & C_{ij}\;.
\end{eqnarray}

Our weights are determined by minimising the stochasticity between the true and reconstructed density fields. 
Defining the stochasticity $E$ as
\begin{equation}
 E^2=1-2\boldsymbol{b}^{T}\boldsymbol{w}+\boldsymbol{w}^{T}\boldsymbol{C}\boldsymbol{w}/P\;,
\end{equation}
the minimum in the stochasticity is reached when the weight is
\begin{equation}\label{eqn:opt_w}
 \boldsymbol{w}_{\rm opt}=(\boldsymbol{C}/P)^{-1}\boldsymbol{b}\;,
\end{equation}
this implies the optimal stochasticity is 
\begin{equation}
 E_{\rm opt}^2=1-\boldsymbol{b}^{T}(\boldsymbol{C}/P)^{-1}\boldsymbol{b}\;.
\end{equation}

The halos are divided into five (logarithmically) equally-spaced mass bins, from the smallest masses (halos with 20 
particles) in bin 1, to bin 5 which includes halos up to 70\% of the highest mass found in the simulation. This most 
massive bin has only nine halos. 
To evaluate the weights, we calculated the power spectrum for each bin and the cross correlation between the bins and 
the total matter field.

In Fig.~\ref{fig:weight} we show the optimal weights to be applied to each mass bin. 
Due to the limited numbers of objects, the weights are very noisy on large scales, particularly for the highest mass 
halos. 
It is easy to understand why the lowest mass bin has a higher normalisation: lower mass halos are less biased and 
better reflect the matter density field not in halos, hence they are up-weighted with respect to the other halos in 
the density reconstruction. This can be seen in both the N-body and the 2LPT simulations. 
On small scales, the amplitude drops a factor of 2.5 going from the lowest mass bin to the highest mass bin in the 
N-body simulation, while for the 2LPT simulation differences in the weights are more pronounced.

\begin{figure}
 \centering
 \includegraphics[width=0.3\textwidth,angle=-90]{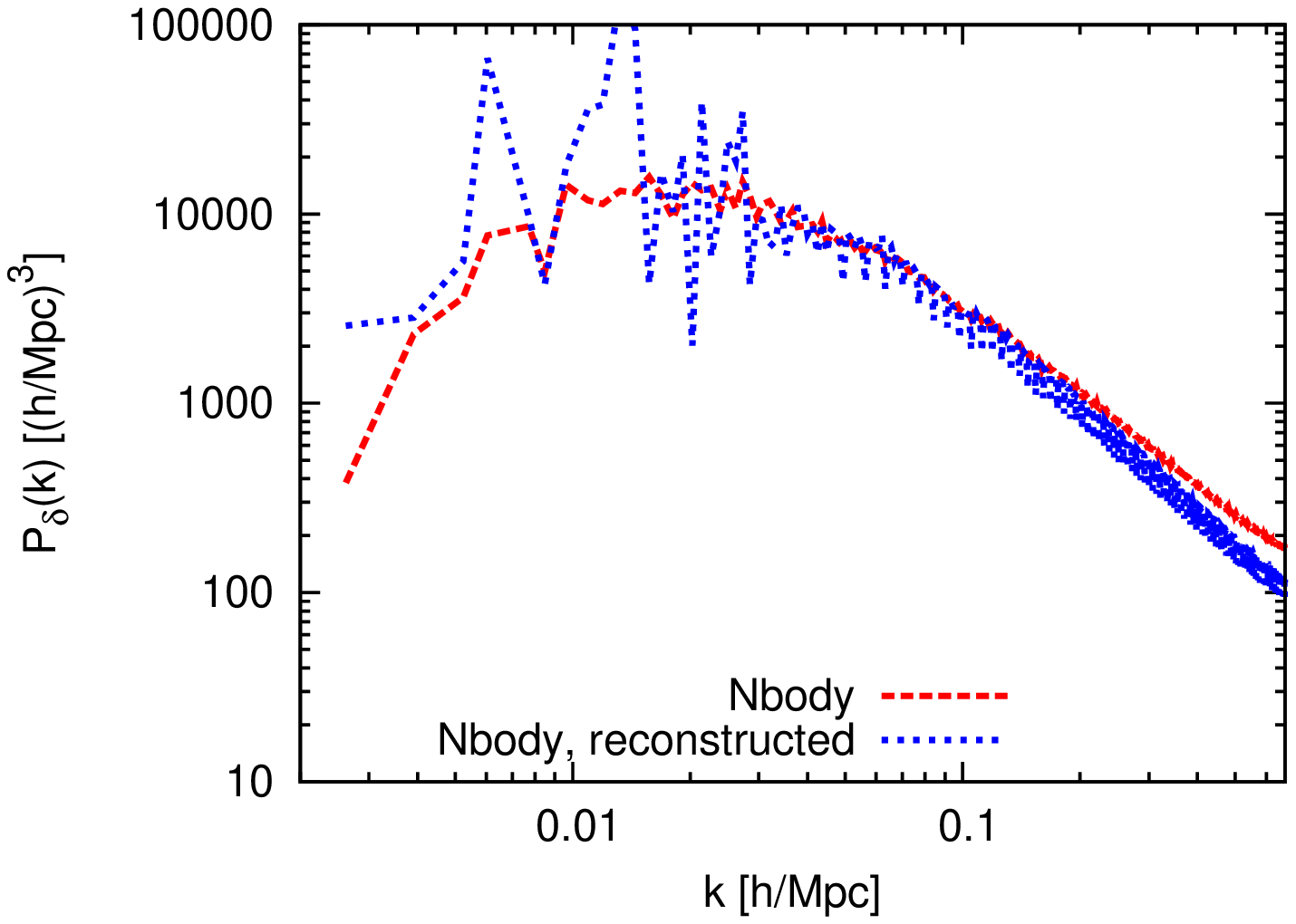}
 \includegraphics[width=0.3\textwidth,angle=-90]{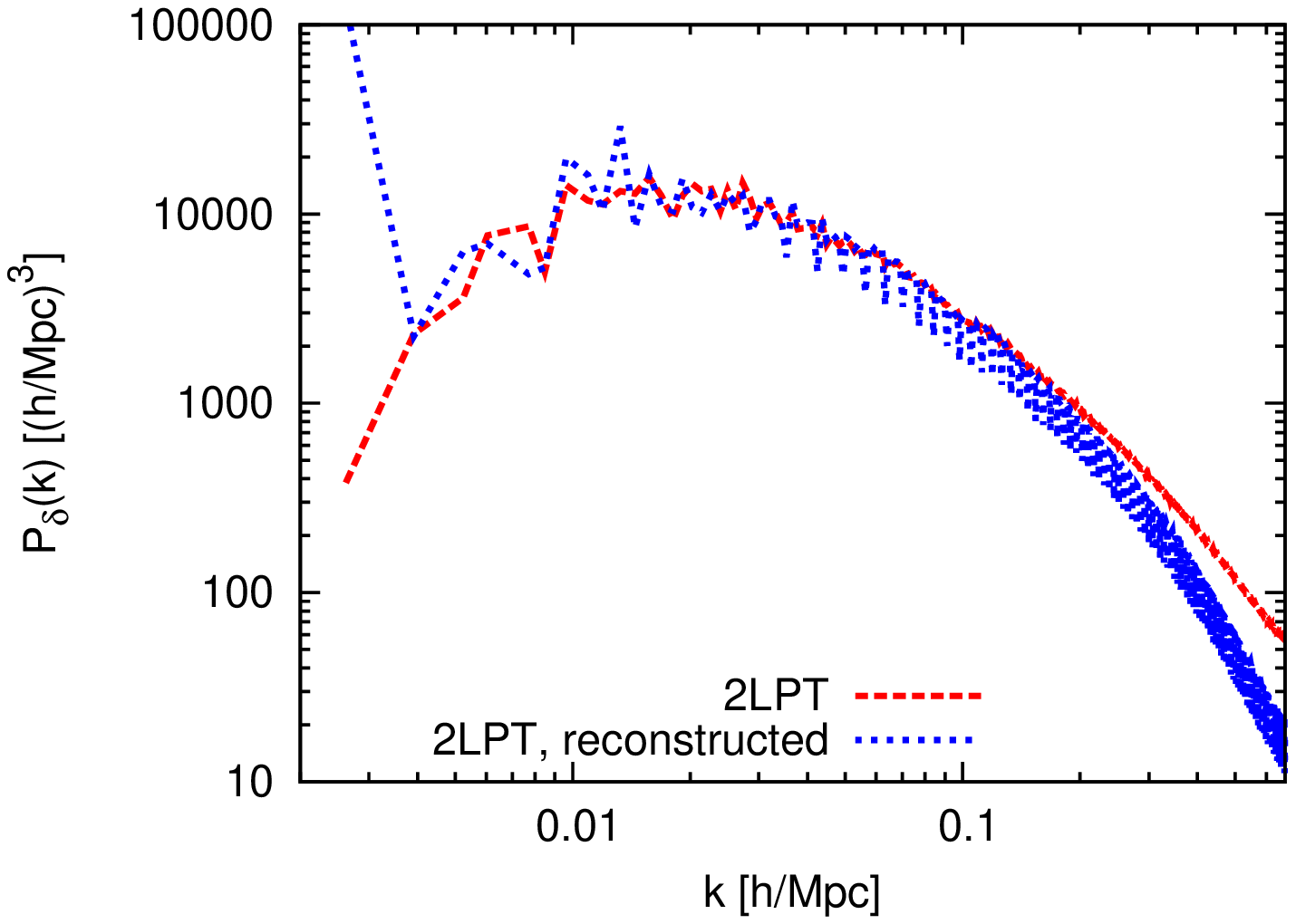}
 \caption{Comparison between the total mass power spectrum (red dashed line) with the reconstructed one using the 
 optimal weights evaluated in Eq.~\ref{eqn:opt_w} (blue short-dashed line). Upper (lower) panel shows results 
 for the N-body (2LPT) simulations.}
 \label{fig:rec3Dspectrum}
\end{figure}

In Fig.~\ref{fig:rec3Dspectrum} we show the reconstructed three dimensional matter power spectrum using the mass 
weighting. 
The reconstruction on large scales is relatively noisy, though the reconstructed matter power spectrum follows the 
expected trend. 
Good agreement is achieved for the mildly non-linear regime, with differences of the order of 10\% for 
$k\approx 0.15$~h/Mpc. 
For smaller scales, deeply in the non-linear regime, we observe a lack of power that becomes more pronounced for 
higher wave numbers. 
The effect of the shot noise is suppressed with the weighting, making the spectrum more reliable over a larger range 
of frequencies. 
In addition, now the spectra agree over a large range of frequencies without any additional rescaling due to the 
different effective matter density parameter, as the weights compensate for this.

In terms of the matter density perturbations, the effective convergence $\kappa(\vec{\theta})$, for sources at a 
fixed redshift, is written as
\begin{equation}
 \kappa(\vec{\theta})=\frac{3\Omega_{\rm m,0}}{2}\left(\frac{H_0}{c}\right)^2
 \int_0^{w}~dw^{\prime}\frac{f_{K}(w^{\prime})f_{K}(w-w^{\prime})}{f_{K}(w)}
 \frac{\delta[f_{K}(w^{\prime}\vec{\theta}),w^{\prime}]}{a}\;,
\end{equation}
where $a$ is the scale factor, $w$ the comoving distance and $f_{K}$ a function that takes into account the 3D 
curvature of the space-time. 
In Fourier space, $\delta$ can be viewed as the weighted sum of the contributions of all the different mass bins (see 
Eq.~\ref{eqn:CBSdelta}). 
Therefore the effective convergence power spectrum is the integration along the line of sight of the reconstructed 
matter power spectrum:
\begin{equation}
 P_{\delta}(k)=\sum_{i=1}^{N_{\rm bins}}w^2_{i}(k)P_{i}(k)+2\sum_{i=1,j>i}^{N_{\rm bins}}w_{i}(k)w_{j}(k)P_{ij}(k)\;,
\end{equation}
where $P_{i}(k)$ represents the matter power spectrum of the $i$-th bin and $P_{ij}(k)$ the cross-correlation between 
the different bins. 
Using these relations, we numerically evaluate the lensing power spectrum from the halo catalogue. 
The matter power spectrum is binned only for a given set of wavelengths, therefore when the evaluation of the matter 
power spectrum has to be done outside this range, we extrapolate the values of the spectrum to both smaller and larger 
scales. 

\begin{figure}
 \centering
 \includegraphics[width=0.3\textwidth,angle=-90]{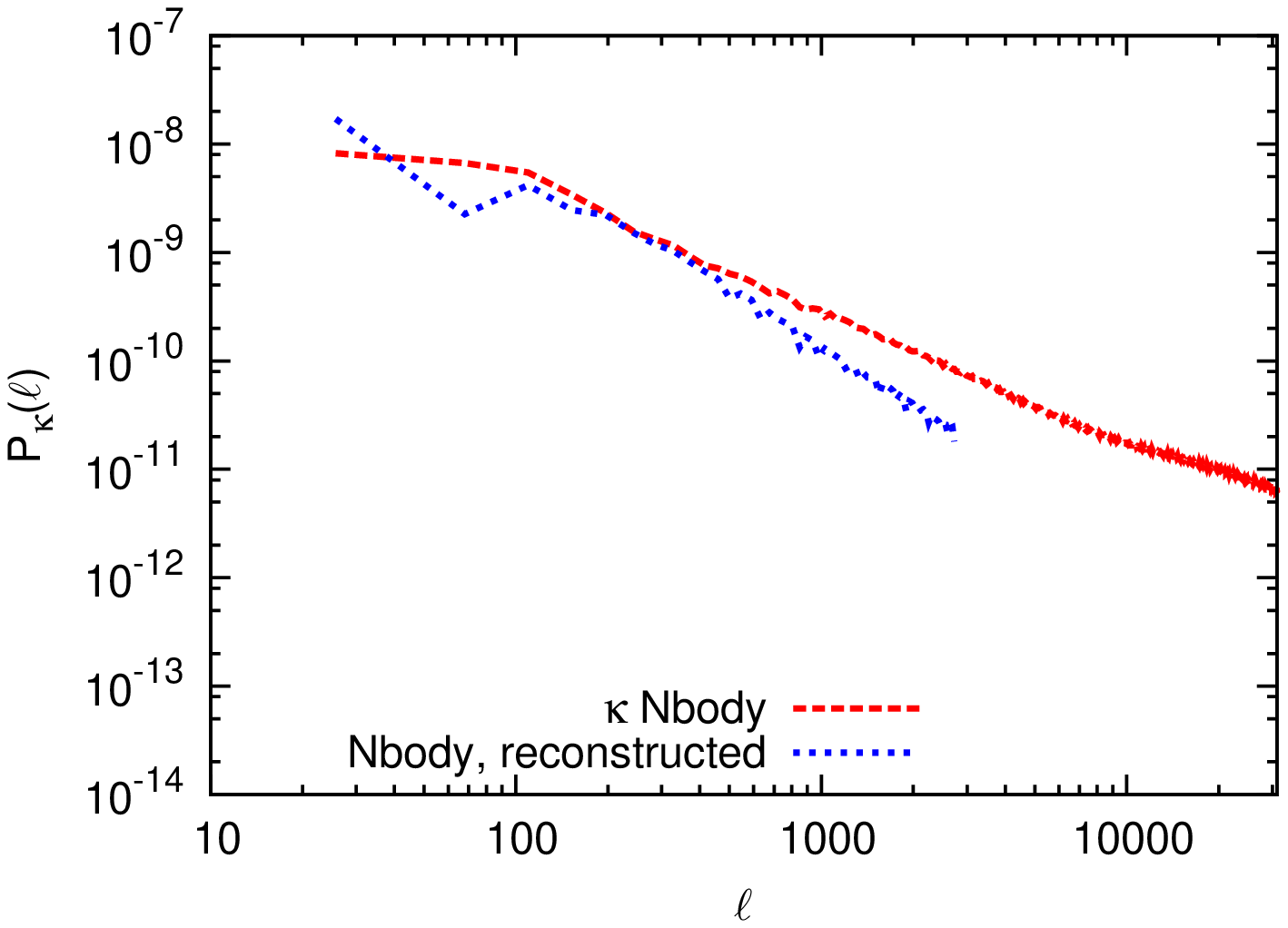}
 \includegraphics[width=0.3\textwidth,angle=-90]{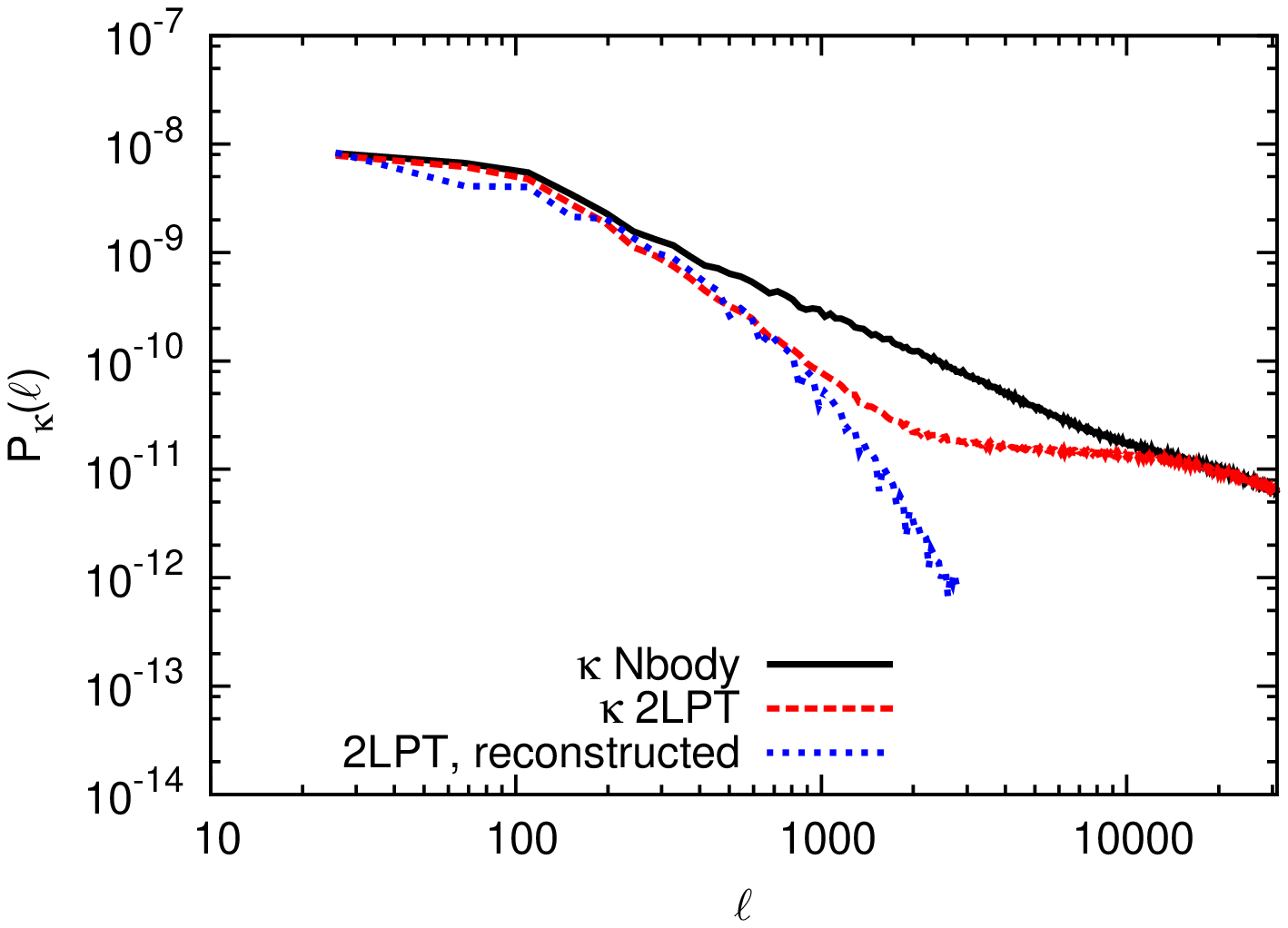}
 \caption{Comparison between the lensing power spectrum obtained from the full the matter distribution (red dashed 
 line) and the lensing power spectrum evaluated from the reconstructed density field (blue short-dashed line). Upper 
 (lower) panel shows results for the N-body (2LPT) simulations.}
 \label{fig:rec2Dspectrum}
\end{figure}

We show our results in Fig.~\ref{fig:rec2Dspectrum}. 
As expected the reconstruction is not perfect and is limited to a limited range of frequencies. 
On large scales the reconstructed lensing power spectrum agrees well with the original one, but at higher wave numbers 
the reconstructed spectrum has a lower amplitude: 
this reflects the lack of power present in the reconstructed three-dimensional matter power spectrum. 
Interestingly, the reconstruction for the 2LPT case works better than for the N-body simulation. 
However, what is reconstructed is the lensing prediction for the 2LPT particles, and so greatly under-produces power 
compared to the N-body lensing prediction, as seen in the bottom panel of Fig.~\ref{fig:rec2Dspectrum}.

\section{Conclusions}\label{sect:conclusions}
Our focus has been to understand how well approximations to large scale structure, in particular those based on the 
halo model, can reproduce the statistics characterising large scale structure observations. Particularly challenging 
in this regard are lensing statistics, where projection effects mean that large and small scale features can jointly 
determine the observations on a given angular scale. Lensing studies are often based on ray-tracing through numerical 
simulations, but these methods are computationally expensive and it is essential to understand which fast methods can 
make reliable lensing predictions.

In this work we have studied whether mock halo catalogues, based either on full simulations or fast 2LPT 
realisations, could be of value for pushing lensing analyses into the non-linear regime. Such halo model approaches, 
if based on simple templates such as the NFW profile, can take advantage of analytical methods to calculate the 
lensing effect and so be much faster. By comparing fully non-linear simulations to mock catalogue methods over a huge 
cosmic volume, we have quantified where the halo model approaches appear to fail.

We have considered a gradual progression from the full N-body simulations to the halo catalogues. The first step 
beyond directly using the N-body simulation was to coherently rotate the halos; and we found that the halo 
orientations did not have a significant impact on either the matter power spectrum or the lensing convergence 
spectrum. However, when the halo particles were rotated incoherently, effectively smearing out the internal halo 
substructures, the matter power spectrum was reduced by 5\% on the smallest scales we probed. The lensing convergence 
was impacted more dramatically, as this blurring of halos resulted in a significant suppression of power even on 
intermediate angular scales (see Fig.~\ref{fig:kappa_rot}).

Directly replacing the halo particles with NFW halos had a similar effect, though in detail it depends on the assumed 
prescription for the concentration evolution. 
We also find that the effect of ignoring substructure is stronger in bigger halos. 
Finally, when we consider the halo catalogues alone, we can still reproduce the density power spectrum on large scales 
with a suitable rescaling, at least up to scales $k\approx 0.1$~h/Mpc, where shot noise in the catalogue begins to 
dominate. 
But this cannot be done for the convergence spectrum; on large scales, the cosmic web of particles not included in 
halos plays an essential role in determining not only the amplitude but also the shape of the lensing spectrum.

One fast means of creating halo catalogues is through a 2LPT approach. 2LPT realisations entirely miss fully 
non-linear structures, and on their own would be inadequate for pushing predictions into the non-linear regime. 
Methods such as PTHalos improve the non-linear properties by identifying embryonic halos in the 2LPT 
simulations and replacing them with more realistic halo masses based on abundance matching to full simulations. 
These can improve agreement in the halo statistics, including the lensing power spectrum.

We also considered an optimal reconstruction of the matter density field by reweighing halos according to their 
mass, following the work by \cite{Cai2011}. 
In this case the reconstructed matter power spectrum agrees reasonably well with the original matter power spectrum 
obtained from the particle distribution and no scaling is necessary to compensate for the different matter density 
parameters. The method worked better on large scales in reproducing the shape of the lensing convergence spectrum, 
compared with the simple halo approach. However, the lack of power on small scales remains an outstanding issue.

We demonstrate that both cosmic web and substructures are important in shaping the total lensing signal. While the 
cosmic web is responsible for the overall shape and contribution on large scales, substructures are responsible for a 
quite substantial fraction of the power on small scales.

To improve the performance of halo methods for lensing on non-linear scales, at a minimum we must include some 
additional substructure beyond the assumption that halos have simple spherically symmetric profiles. An obvious 
possibility to explore is the impact of the halo ellipticity; it would be interesting to understand whether 
extracting and modelling the halo ellipticities could help address the power deficits we have seen in the lensing 
spectrum. 
Another possibility is to include substructures, such as satellite galaxies, following, for example, the work of 
\cite{Giocoli2008,Giocoli2010} in characterising the number and masses of satellite halos. 
Alternatively, one could create a library of realistic halos (or their associated convergence maps) that could be 
pasted randomly onto halo catalogues.

\section*{Acknowledgements}
The authors thank Roman Scoccimarro and the Las Damas collaboration to provide the N-body simulations we used in 
this work. The analysis was performed on the Intel SCIAMA High Performance Compute (HPC) cluster which is supported 
by the ICG, SEPNet and the University of Portsmouth. F.~P., D.~B. and R.~C. are supported by STFC grant ST/H002774/1. 
W.~P. is grateful for support from the UK Science and Technology Facilities Research Council through the grant 
ST/I001204/1, and the European Research Council through the grant "Darksurvey". MM acknowledges support from the 
European Research Council through the grant ”TESTDE”.

\bibliographystyle{mn2e}
\bibliography{MockCatalogues.bbl}

\label{lastpage}

\end{document}